\documentclass{nature}

\usepackage{graphicx} % Required for the inclusion of images
\usepackage{lineno}
\usepackage{tabularx, booktabs}
\usepackage{adjustbox}
\usepackage{longtable}
\usepackage{ltablex}
\usepackage{threeparttablex}

\makeatletter
\def\makeLineNumberLeft{%
  \linenumberfont\llap{\hb@xt@\linenumberwidth{\LineNumber\hss}\hskip\linenumbersep}% left line number
  \hskip\columnwidth% skip over column of text
  \rlap{\hskip\linenumbersep\hb@xt@\linenumberwidth{\hss\LineNumber}}\hss}% right line number
\leftlinenumbers% Re-issue [left] option
\makeatother

\def\gtrsim{\mathrel{\hbox{\rlap{\hbox{\lower5pt\hbox{$\sim$}}}\hbox{$>$}}}}
\newcommand{\jj}[2]{\mbox{$J = #1\rightarrow#2$}}

\newcommand{\arcsec}{\mbox{$^{\prime \prime}$}}
\newcommand{\degree}{\mbox{$^\circ$}}
\newcommand{\rsquo}{{'}}
\newcommand{\mdash}{\mbox{$-$}}

\newcolumntype{Y}{>{\centering\arraybackslash}X}

\newcommand{\nat}{{ Nature }}
\newcommand{\aap}{{Astron.~Astrophys.}}
\newcommand{\aj}{{ Astron.~J. }}
\newcommand{\apj}{{ Astrophys.~J. }}
\newcommand{\araa}{{Ann.~Rev.~Astron.~Astrophys.}}
\newcommand{\apjl}{{Astrophys.~J.~Lett. }}
\newcommand{\apjs}{{Astrophys.~J.~Suppl. }}
\newcommand{\mnras}{{Mon. Not. R. Astron. Soc. }}
\newcommand{\jqsrt}{{J.~Quant.~Spectrosc.~Ra.}}
\let\oldthebibliography=\thebibliography
\let\oldendthebibliography=\endthebibliography
\renewenvironment{thebibliography}[1]{%
     \oldthebibliography{#1}%
     \setcounter{enumiv}{30}%
}{\oldendthebibliography}

\title{The ice composition in the disk around V883 Ori revealed by its stellar outburst}

\author{Jeong-Eun Lee$^{1}$, Seokho Lee$^{1}$, Giseon Baek$^{1}$, Yuri Aikawa$^{2}$, Lucas Cieza$^{3}$, Sung-Yong Yoon$^{1}$, 
        Gregory Herczeg$^{4}$, Doug Johnstone$^{5}$, Simon Casassus$^6$}

\begin{document}

\maketitle

\begin{affiliations}
 \item School of Space Research, Kyung Hee University, 1732, Deogyeong-Daero, Giheung-gu, Yongin-shi, Gyunggi-do 17104, Korea
 \item Department of Astronomy, University of Tokyo, 7-3-1 Hongo, Bunkyo-ku, Tokyo 113-0033, Japan
% \item Millenium Nucleus 'Protoplanetary discs in ALMA Early Science', Av. Ejercito 441, 8370191, Santiago, RM, Chile
 \item Facultad de Ingenier\'ia y Ciencias, N\'ucleo de Astronom\'ia, Universidad Diego Portales, Av. Ejercito 441. Santiago, Chile	 
 \item Kavli Institute for Astronomy and Astrophysics, Peking University, Yi He Yuan Lu 5, Haidian Qu, 100871, Beijing, PR China
 %\item NRC Herzberg Astronomy and Astrophysics, 5071 West Saanich Road, Victoria, BC, V9E 2E7, Canada
 \item  NRC Herzberg Astronomy and Astrophysics, 5071 West Saanich Road, Victoria, BC, V9E 2E7, Canada 
 %\item  Department of Physics and Astronomy, University of Victoria, Victoria, BC, V8P 5C2, Canada 
 \item Departamento de Astronom\'ia, Universidad de Chile, Casilla 36-D Santiago, Chile
 \end{affiliations}

%\pagestyle{empty}
%\linenumbers

\begin{abstract}
%Earth-like planets form mostly from dry refractory materials in the inner regions of protoplanetary disks; however, they might become habitable if water and organic molecules are delivered to their surfaces and atmospheres by planetesimals formed beyond the sublimation front of water\cite{Chyba1990}. 
Complex organic molecules (COMs), which are the seeds of prebiotic material and precursors of amino acids and sugars, form in the icy mantles of circumstellar dust grains\cite{Herbst2009} but cannot be detected remotely unless they are heated and released to the gas phase.  Around solar-mass stars, water and COMs only sublimate in the inner few au of the disk\cite{dalessio2001}, making them extremely difficult to spatially resolve and study. Sudden increases in the luminosity of the central star will quickly expand the sublimation front (so-called {\it snow line}) to larger radii, as seen previously in the FU Ori outburst of the young star V883 Ori\cite{Cieza2016}.  In this paper, we take advantage of the rapid increase in disk temperature of V883 Ori to detect and analyze five different COMs, methanol, acetone, acetonitrile, acetaldehyde, and methyl formate, in spatially-resolved submillimeter observations.
The COMs abundances in V883 Ori is in reasonable agreement with cometary values\cite{LeRoy2015}.
This result suggests that outbursting young stars can provide a special opportunity to study the ice composition of material directly related to planet formation.

\end{abstract}

Observations of comets and asteroids show that the Solar nebula was rich in 
water and organic molecules. The inventory from the recent Rosetta mission 
to comet 67P/Churyumov–Gerasimenko includes many complex organic molecules (COMs)\cite{Altwegg2017}
as well as prebiotic molecules.
 These organics, together with water, could have been brought to the young Earth's surface by comets\cite{Chyba1990}. 
 Since the ice composition of comets is similar to ices in molecular clouds, it has long been debated if 
cometary ices originate in the interstellar matter (ISM). The evolution of volatiles during planet formation may be traced by comparing the abundances from the youngest phases of star and disk formation, the protostellar core phase (when the star and disk are still growing from a circumstellar envelope), to the older phases of disk evolution.
The Atacama Large Millimeter/submillimeter  Array (ALMA) revealed that in some protostellar cores, called hot corinos, the warm gas contains various COMs that are also detected in comets\cite{Imai2016, Jorgensen2016}.  The variation of chemical composition among comets\cite{LeRoy2015}, however, suggests that chemical processes could be active even after  material from the ISM is incorporated to the disk\cite{Mumma2011}.  The chemical reactions, in turn, depend on various physical parameters in the disk, e.g. ionization rate and dust-gas decoupling, which are under debate\cite{Furuya2014,Schwarz2018}.

 Despite efforts to observe ices in disks, clear detection of ices in the disk midplane is challenging and requires sophisticated analysis that considers the inclination angle and foreground contamination\cite{pontoppidan2005}. An alternative approach is to observe COMs in the gas phase inside the snow line, defined as the radius outside of which a molecule is predominantly in ice form.  
The location of each snow line depends on the heating of the disk by the central star and by the release of gravitational energy via mass accretion, as well as on the sublimation temperature of the molecule\cite{Hama2013}.  However, for a typical disk around a solar-mass star and a mass accretion rate of $\sim10^{-8}$ M$_\odot$ yr$^{-1}$, the COM and water snow lines are located at a radius of only a few AU\cite{dalessio2001}, too small to be spatially resolved even with ALMA. %and may not be relevant to ices in comets, which form at larger radii.
So far, three gaseous COMs, CH$_3$OH, CH$_3$CN, and HCOOH, have been detected in  non-bursting protoplanetary disks\cite{Walsh2016, Oberg2015, Loomis2018, Bergner2018, Favre2018}, with emission that traces non-thermally desorbed molecules in the outer ($>$10 AU) disk surface, far beyond the snow line.  
Derivation of solid-phase
abundances from those observations is not straightforward; it needs detailed understanding of the non-thermal desorption mechanisms and of gas-phase reactions that could change the abundances after desorption\cite{Bertin2016}.

The abundances of COMs at the disk locations where icy planetesimals form  may instead be best measured during a protostellar outburst\cite{Molyarova2018}, which viscously heats the disk and extends snow line to a much larger radius.  The gas emission from the freshly sublimated COMs should directly trace the abundances of COMs that were previously in the ice phase.
FUors are young stellar objects that exhibit large-amplitude 
outbursts in the optical (change in magnitude $\Delta m_{V} > 4$ mag) and undergo rapid increases of 
accretion rate, by $\sim$3 orders of magnitude (from $10^{-7}$ to $10^{-4}$ M$_\odot$ yr$^{-1}$)\cite{audard2014}. 
The high luminosity due to the enhanced accretion rate shifts the snow line to 
much larger radii\cite{harsono2015}.  For the FUor outburst V883 Ori studied here, the radius of the water snow line at the disk midplane is estimated to be $\sim$40 AU from 1.3 mm continuum emission\cite{Cieza2016}; the snow line at the disk surface may extend to $\sim$160 AU.
 Theoretical studies show that the sublimates are destroyed by the gas-phase reactions 
in several 10$^4$ yr\cite{Nomura2009}. Since the duration of a typical outburst is 
usually $\le$100 years, FUors provide a unique and direct probe to study the composition of fresh sublimates of COMs located in the disk midplane, which were in ice form in the pre-burst phase.
 When the outburst stops, the dust grains will cool instantaneously, so these COMs will be frozen back onto grain surfaces within one year (see Methods).

We observed V883 Ori, a FU Ori star with mass of 1.2 M$_\odot$\cite{Cieza2016} (see Methods), with ALMA in Band 7 on 8 September 2017 with a resolution of $\sim 0.03\arcsec$, designed for continuum images, and on 21 February 2018 with a resolution 
of $\sim 0.2\arcsec$, designed to measure COM emission. 
 From the lower-resolution observation, we detected many COMs lines, including CH$_3$CHO, CH$_3$OCHO, and CH$_3$COCH$_3$, which are robustly detected for the first time in protoplanetary disks (Figure 1).
 CH$_3$CHO and CH$_3$OCHO were detected contemporaneously and independently in the same target\cite{vantHoff2018}. 
 Figure 2 shows the integrated intensity images (contours) and intensity weighted velocity images (colors) of emission coadded in many lines of individual COMs from the lower-resolution observation.
 All COM emission is confined within $\sim$0.4\arcsec, but the emission peaks are offset from the continuum peak.
We find excitation temperatures for $^{13}$CH$_3$OH of 91.5 $\pm$ 3.5 K and for  CH$_3$OCHO of 103.3 $\pm$ 34.9 K (see Supplementary Figure 1). 
 The lines of these two molecules are reasonably optically thin (line optical depth $\le 1$, see Methods).

 While the high resolution data was obtained to image the dust continuum, it also contains some COMs emissions.
The high-resolution images (Figure 3 and Supplementary Figures 2 and 3) reveal that the COM emission originates in a ring region around the water snow line that was previously identified from the dust continuum observations. The $^{13}$CH$_3$OH emission arises only between 0.1$\arcsec$-0.2\arcsec, in contrast to the expectation that COMs emission should be bright inside the snow line.
The  position-velocity map 
presented in Figure 3(c) is reproduced well by ring within a Keplerian disk, 
consistent with the CO emission\cite{Cieza2016}.  

Disks are heated by stellar irradiation and viscous accretion\cite{dalessio2001,Qi2006}. When stellar irradiation dominates, then the disk surface is warmer than the midplane, while the midplane could be warmer when the accretion heating dominates. The location where the water sublimates thus varies both with the distance from the star and
with the height from the midplane, which together defines the snow surface (Supplementary Figure 5). Because the molecular line opacity tends to be higher than that of the dust continuum, the 2-D (radial and vertical) temperature distribution is needed to analyze molecular line emission. 

Since heating by irradiation and accretion are both important for FUors\cite{Dullemond2007}, we tested two models for the optically thin $^{13}$CH$_3$OH line emission: Model A for the disk heated only by irradiation and Model B for the disk heated by an enhanced accretion in the midplane as well as irradiation (see Methods and Supplementary Figure 4 for the details of models).  
Figure 4 compares the results of
the two models with the observed emission distribution of $^{13}$CH$_3$OH $12_{11}-12_{12}$. 
The observed emission decreases inside of its snow line, which is well reproduced by Model A and is a consequence of increasing dust continuum optical depths (see the bottom panels of Figure 4 and Supplementary Figures 4) 
caused by the evaporation of cm-sized icy pebbles (containing micron-sized dust grains)  and the re-coagulation of the bare silicates into submm-sized grains\cite{Schoonenberg2017}.
In Model B, the $^{13}$CH$_3$OH emission declines even more steeply inwards due to the self-absorption caused by the negative temperature gradient in the vertical direction.
Observations with higher spatial resolution and better sensitivity are needed to discriminate between these two models.  Figure 4 also compares the total amount of $^{13}$CH$_3$OH emission with that produced in the disk surface.  The fraction of emission from the midplane reaches a maximum near the methanol snow line, confirming that the COM compositions in the gas phase originate in the ice compositions in the midplane regions that are likely forming planetesimals.
 
The average column densities of COMs over their emitting area are derived by fitting the observed spectra (Figure 1) with XCLASS\cite{Moller2017} (Supplementary Table 1). 
 The abundances with respect to molecular hydrogen are $10^3-10^5$ times higher than those derived from the spatially-extended COMs emission of non-bursting disks ($\sim 10^{-13}-10^{-11}$)\cite{Walsh2016, Oberg2015, Loomis2018, Bergner2018, Favre2018}, confirming that our observations probe the thermal sublimation of COMs. 
 The D/H and $^{13}$C/$^{12}$C ratios of CH$_3$OH are 0.16 and 0.13, which are significantly higher than the ratios in protostellar cores and elemental abundances.
 While the high D/H ratio could be due to deuterium fractionation in the dense and cold grain surfaces\cite{ Hama2013}, the high $^{13}$C/$^{12}$C ratio indicates that the column density of CH$_3$OH is underestimated due to the high optical depth of the lines, requiring a more detailed treatment than that taken by XCLASS.  We thus assume that the actual column density of CH$_3$OH is equal to that of $^{13}$CH$_3$OH multiplied by the elemental abundance of $^{12}$C/$^{13}$C$=60$ (see Methods).

 We find most COMs abundances relative to CH$_3$OH, i.e. their column density ratios, are higher than those in the hot corino IRAS 16293 B\cite{Jorgensen2016, Lykke2017, Calcutt2018b}; the abundances of CH$_3$COCH$_3$, CH$_3$CHO, and CH$_3$OCHO are higher than those in IRAS 16293 B by factors of 4 to 16 (Supplementary Table 1). The abundances of CH$_3$COCH$_3$ and CH$_3$CHO in both V883 Ori and IRAS 16293 B agree with those in comet 67P/Churyumov–Gerasimenko\cite{LeRoy2015, Mumma2011} by a factor of a few. 
 Acetone is not detected by by the Rosetta Orbiter Spectrometer
for Ion and Neutral Analysis (ROSINA) on board Rosetta but was estimated to be as abundant as CH$_3$CHO in 67P/Churyumov–Gerasimenko by the
Cometary Sampling and Composition Experiment (COSAC)\cite{goesmann2015}, which could be in reasonable agreement with V883 Ori. CH$_3$CN abundance in V883 Ori, on the other hand, is lower than in IRAS 
16293 B (comets) by a factor of 3 (an order of magnitude). The CH$_3$CN/CH$_3$OH abundance ratio is also much lower than that in cold vapor in outer radii of other disks ($\sim$1)\cite{Loomis2018, Bergner2018}. These abundances indicate that formation and destruction of COMs 
continue after the volatiles are incorporated to the disk.

Our observations and analysis of V883 Ori demonstrate that observations of FUors are unique in revealing fresh sublimates and thus ice composition in protoplanetary disks.  Although FUors are rare, they span a range of evolutionary states: some outbursts occur while the system is still being fed by their envelope, other outbursts occur in systems that have only disks and no envelope\cite{audard2014}.  
This diversity may provide us with the leverage to investigate how the chemical evolution of complex molecules in protoplanetary disks leaves its imprints on the products of planet formation.

\clearpage

\begin{oldthebibliography}{10}
%{{{1
\expandafter\ifx\csname url\endcsname\relax
  \def\url#1{\texttt{#1}}\fi
\expandafter\ifx\csname urlprefix\endcsname\relax\def\urlprefix{URL }\fi
\providecommand{\bibinfo}[2]{#2}
\providecommand{\eprint}[2][]{\url{#2}}

\bibitem{Herbst2009}
\bibinfo{author}{{Herbst}, E.} \& \bibinfo{author}{{van Dishoeck}, E.~F.}
\newblock \bibinfo{title}{{Complex Organic Interstellar Molecules}}.
\newblock \emph{\bibinfo{journal}{\araa}} \textbf{\bibinfo{volume}{47}},
  \bibinfo{pages}{427--480} (\bibinfo{year}{2009}).

\bibitem{dalessio2001}
\bibinfo{author}{{D'Alessio}, P.}, \bibinfo{author}{{Calvet}, N.} \&
  \bibinfo{author}{{Hartmann}, L.}
\newblock \bibinfo{title}{{Accretion Disks around Young Objects. III. Grain
  Growth}}.
\newblock \emph{\bibinfo{journal}{\apj}} \textbf{\bibinfo{volume}{553}},
  \bibinfo{pages}{321--334} (\bibinfo{year}{2001}).
\newblock \eprint{astro-ph/0101443}.

\bibitem{Cieza2016}
\bibinfo{author}{{Cieza}, L.~A.} \emph{et~al.}
\newblock \bibinfo{title}{{Imaging the water snow-line during a protostellar
  outburst}}.
\newblock \emph{\bibinfo{journal}{\nat}} \textbf{\bibinfo{volume}{535}},
  \bibinfo{pages}{258--261} (\bibinfo{year}{2016}).
\newblock \eprint{1607.03757}.

\bibitem{LeRoy2015}
\bibinfo{author}{{Le Roy}, L.} \emph{et~al.}
\newblock \bibinfo{title}{{Inventory of the volatiles on comet
  67P/Churyumov-Gerasimenko from Rosetta/ROSINA}}.
\newblock \emph{\bibinfo{journal}{\aap}} \textbf{\bibinfo{volume}{583}},
  \bibinfo{pages}{A1} (\bibinfo{year}{2015}).

%\bibitem{Crovisier2006}
%\bibinfo{author}{{Crovisier}, J.}
%\newblock \bibinfo{title}{{New trends in cometary chemistry}}.
%\newblock \emph{\bibinfo{journal}{Faraday Discussions}}
%  \textbf{\bibinfo{volume}{133}}, \bibinfo{pages}{375} (\bibinfo{year}{2006}).

\bibitem{Altwegg2017}
\bibinfo{author}{{Altwegg}, K.} \emph{et~al.}
\newblock \bibinfo{title}{{Organics in comet 67P - a first comparative analysis
  of mass spectra from ROSINA-DFMS, COSAC and Ptolemy}}.
\newblock \emph{\bibinfo{journal}{\mnras}} \textbf{\bibinfo{volume}{469}},
  \bibinfo{pages}{S130--S141} (\bibinfo{year}{2017}).

\bibitem{Chyba1990}
\bibinfo{author}{{Chyba}, C.~F.}, \bibinfo{author}{{Thomas}, P.~J.},
  \bibinfo{author}{{Brookshaw}, L.} \& \bibinfo{author}{{Sagan}, C.}
\newblock \bibinfo{title}{{Cometary Delivery of Organic Molecules to the Early
  Earth}}.
\newblock \emph{\bibinfo{journal}{Science}} \textbf{\bibinfo{volume}{249}},
 \bibinfo{pages}{366--373} (\bibinfo{year}{1990}).
 
\bibitem{Imai2016}
\bibinfo{author}{{Imai}, M.} \emph{et~al.}
\newblock \bibinfo{title}{{Discovery of a Hot Corino in the Bok Globule B335}}.
\newblock \emph{\bibinfo{journal}{\apjl}} \textbf{\bibinfo{volume}{830}},
  \bibinfo{pages}{L37} (\bibinfo{year}{2016}).
\newblock \eprint{1610.03942}.

\bibitem{Jorgensen2016}
\bibinfo{author}{{J{\o}rgensen}, J.~K.} \emph{et~al.}
\newblock \bibinfo{title}{{The ALMA Protostellar Interferometric Line Survey
  (PILS). First results from an unbiased submillimeter wavelength line survey
  of the Class 0 protostellar binary IRAS 16293-2422 with ALMA}}.
\newblock \emph{\bibinfo{journal}{\aap}} \textbf{\bibinfo{volume}{595}},
  \bibinfo{pages}{A117} (\bibinfo{year}{2016}).
\newblock \eprint{1607.08733}.

\bibitem{Mumma2011}
\bibinfo{author}{{Mumma}, M.~J.} \& \bibinfo{author}{{Charnley}, S.~B.}
\newblock \bibinfo{title}{{The Chemical Composition of Comets{\mdash}Emerging
  Taxonomies and Natal Heritage}}.
\newblock \emph{\bibinfo{journal}{\araa}} \textbf{\bibinfo{volume}{49}},
  \bibinfo{pages}{471--524} (\bibinfo{year}{2011}).

\bibitem{Furuya2014}
\bibinfo{author}{{Furuya}, K.} \& \bibinfo{author}{{Aikawa}, Y.}
\newblock \bibinfo{title}{{Reprocessing of Ices in Turbulent Protoplanetary
  Disks: Carbon and Nitrogen Chemistry}}.
\newblock \emph{\bibinfo{journal}{\apj}} \textbf{\bibinfo{volume}{790}},
  \bibinfo{pages}{97} (\bibinfo{year}{2014}).
\newblock \eprint{1406.3507}.

\bibitem{Schwarz2018}
\bibinfo{author}{{Schwarz}, K.~R.} \emph{et~al.}
\newblock \bibinfo{title}{{Unlocking CO Depletion in Protoplanetary Disks. I.
  The Warm Molecular Layer}}.
\newblock \emph{\bibinfo{journal}{\apj}} \textbf{\bibinfo{volume}{856}},
  \bibinfo{pages}{85} (\bibinfo{year}{2018}).
\newblock \eprint{1802.02590}.

\bibitem{pontoppidan2005}
\bibinfo{author}{{Pontoppidan}, K.~M.} \emph{et~al.}
\newblock \bibinfo{title}{{Ices in the Edge-on Disk CRBR 2422.8-3423: Spitzer
  Spectroscopy and Monte Carlo Radiative Transfer Modeling}}.
\newblock \emph{\bibinfo{journal}{\apj}} \textbf{\bibinfo{volume}{622}},
  \bibinfo{pages}{463--481} (\bibinfo{year}{2005}).
\newblock \eprint{astro-ph/0411367}.

\bibitem{Hama2013}
\bibinfo{author}{{Hama}, T.} \& \bibinfo{author}{{Watanabe}, N.}
\newblock \bibinfo{title}{{Surface Processes on Interstellar Amorphous Solid
  Water: Adsorption, Diffusion, Tunneling Reactions, and Nuclear-Spin
  Conversion}}.
%\newblock \emph{\bibinfo{journal}{Chemical Reviews}}
\newblock \emph{\bibinfo{journal}{Chem.~Rev.}}
  \textbf{\bibinfo{volume}{113}}, \bibinfo{pages}{8783--8839}
  (\bibinfo{year}{2013}).

\bibitem{Walsh2016}
\bibinfo{author}{{Walsh}, C.} \emph{et~al.}
\newblock \bibinfo{title}{{First Detection of Gas-phase Methanol in a
  Protoplanetary Disk}}.
\newblock \emph{\bibinfo{journal}{\apjl}} \textbf{\bibinfo{volume}{823}},
  \bibinfo{pages}{L10} (\bibinfo{year}{2016}).
\newblock \eprint{1606.06492}.

\bibitem{Oberg2015}
  \b ibinfo{author}{{{\"O}berg}, K.~I.} \emph{et~al.}
\newblock \bibinfo{title}{{The comet-like composition of a protoplanetary disk
  as revealed by complex cyanides}}.
\newblock \emph{\bibinfo{journal}{\nat}} \textbf{\bibinfo{volume}{520}},
  \bibinfo{pages}{198--201} (\bibinfo{year}{2015}).
\newblock \eprint{1505.06347}.

\bibitem{Loomis2018}
\bibinfo{author}{{Loomis}, R.~A.} \emph{et~al.}
\newblock \bibinfo{title}{{Detecting Weak Spectral Lines in Interferometric
  Data through Matched Filtering}}.
\newblock \emph{\bibinfo{journal}{\aj}} \textbf{\bibinfo{volume}{155}},
  \bibinfo{pages}{182} (\bibinfo{year}{2018}).
\newblock \eprint{1803.04987}.

\bibitem{Bergner2018}
\bibinfo{author}{{Bergner}, J.~B.}, \bibinfo{author}{{Guzm{\'a}n}, V.~G.},
    \ bibinfo{author}{{{\"O}berg}, K.~I.}, \bibinfo{author}{{Loomis}, R.~A.} \&
  \bibinfo{author}{{Pegues}, J.}
\newblock \bibinfo{title}{{A Survey of CH$_{3}$CN and HC$_{3}$N in
  Protoplanetary Disks}}.
\newblock \emph{\bibinfo{journal}{\apj}} \textbf{\bibinfo{volume}{857}},
  \bibinfo{pages}{69} (\bibinfo{year}{2018}).
\newblock \eprint{1803.04986}.

\bibitem{Favre2018}
\bibinfo{author}{{Favre}, C.} \emph{et~al.}
\newblock \bibinfo{title}{{First Detection of the Simplest Organic Acid in a
  Protoplanetary Disk}}.
\newblock \emph{\bibinfo{journal}{\apjl}} \textbf{\bibinfo{volume}{862}},
  \bibinfo{pages}{L2} (\bibinfo{year}{2018}).
\newblock \eprint{1807.05768}.

\bibitem{Bertin2016}
\bibinfo{author}{{Bertin}, M.} \emph{et~al.}
\newblock \bibinfo{title}{{UV Photodesorption of Methanol in Pure and CO-rich
  Ices: Desorption Rates of the Intact Molecule and of the Photofragments}}.
\newblock \emph{\bibinfo{journal}{\apjl}} \textbf{\bibinfo{volume}{817}},
  \bibinfo{pages}{L12} (\bibinfo{year}{2016}).
\newblock \eprint{1601.07027}.

\bibitem{Molyarova2018}
\bibinfo{author}{{Molyarova}, T.} \emph{et~al.}
\newblock \bibinfo{title}{{Chemical Signatures of the FU Ori Outbursts}}.
\newblock \emph{\bibinfo{journal}{\apj}} \textbf{\bibinfo{volume}{866}},
  \bibinfo{pages}{46} (\bibinfo{year}{2018}).
%\newblock \emph{\bibinfo{journal}{ArXiv e-prints}}  (\bibinfo{year}{2018}).
\newblock \eprint{1809.01925}.

\bibitem{audard2014}
\bibinfo{author}{{Audard}, M.} \emph{et~al.}
\newblock \bibinfo{title}{{Episodic Accretion in Young Stars}}.
\newblock \emph{\bibinfo{journal}{Protostars and Planets VI}}
  \bibinfo{pages}{387--410} (\bibinfo{year}{2014}).
\newblock \eprint{1401.3368}.

\bibitem{harsono2015}
\bibinfo{author}{{Harsono}, D.}, \bibinfo{author}{{Bruderer}, S.} \&
  \bibinfo{author}{{van Dishoeck}, E.~F.}
\newblock \bibinfo{title}{{Volatile snowlines in embedded disks around low-mass
  protostars}}.
\newblock \emph{\bibinfo{journal}{\aap}} \textbf{\bibinfo{volume}{582}},
  \bibinfo{pages}{A41} (\bibinfo{year}{2015}).
\newblock \eprint{1507.07480}.

\bibitem{Nomura2009}
\bibinfo{author}{{Nomura}, H.}, \bibinfo{author}{{Aikawa}, Y.},
  \bibinfo{author}{{Nakagawa}, Y.} \& \bibinfo{author}{{Millar}, T.~J.}
\newblock \bibinfo{title}{{Effects of accretion flow on the chemical structure
  in the inner regions of protoplanetary disks}}.
\newblock \emph{\bibinfo{journal}{\aap}} \textbf{\bibinfo{volume}{495}},
  \bibinfo{pages}{183--188} (\bibinfo{year}{2009}).
\newblock \eprint{0810.4610}.

\bibitem{vantHoff2018}
\bibinfo{author}{{van {\rsquo}t Hoff}, M.~L.~R.} \emph{et~al.}
\newblock \bibinfo{title}{{Methanol and its Relation to the Water Snowline in
  the Disk around the Young Outbursting Star V883 Ori}}.
\newblock \emph{\bibinfo{journal}{\apjl}} \textbf{\bibinfo{volume}{864}},
  \bibinfo{pages}{L23} (\bibinfo{year}{2018}).
\newblock \eprint{1808.08258}.

\bibitem{Qi2006}
\bibinfo{author}{{Qi}, C.} \emph{et~al.}
\newblock \bibinfo{title}{{CO J = 6-5 Observations of TW Hydrae with the
  Submillimeter Array}}.
\newblock \emph{\bibinfo{journal}{\apjl}} \textbf{\bibinfo{volume}{636}},
  \bibinfo{pages}{L157--L160} (\bibinfo{year}{2006}).
\newblock \eprint{astro-ph/0512122}.

\bibitem{Dullemond2007}
\bibinfo{author}{{Dullemond}, C.~P.}, \bibinfo{author}{{Hollenbach}, D.},
  \bibinfo{author}{{Kamp}, I.} \& \bibinfo{author}{{D'Alessio}, P.}
\newblock \bibinfo{title}{{Models of the Structure and Evolution of
  Protoplanetary Disks}}.
\newblock \emph{\bibinfo{journal}{Protostars and Planets V}}
  \bibinfo{pages}{555--572} (\bibinfo{year}{2007}).
\newblock \eprint{astro-ph/0602619}.

\bibitem{Schoonenberg2017}
\bibinfo{author}{{Schoonenberg}, D.}, \bibinfo{author}{{Okuzumi}, S.} \&
  \bibinfo{author}{{Ormel}, C.~W.}
\newblock \bibinfo{title}{{What pebbles are made of: Interpretation of the V883
  Ori disk}}.
\newblock \emph{\bibinfo{journal}{\aap}} \textbf{\bibinfo{volume}{605}},
  \bibinfo{pages}{L2} (\bibinfo{year}{2017}).
\newblock \eprint{1708.03328}.

\bibitem{Moller2017}
\bibinfo{author}{{M{\"o}ller}, T.}, \bibinfo{author}{{Endres}, C.} \&
  \bibinfo{author}{{Schilke}, P.}
\newblock \bibinfo{title}{{eXtended CASA Line Analysis Software Suite
  (XCLASS)}}.
\newblock \emph{\bibinfo{journal}{\aap}} \textbf{\bibinfo{volume}{598}},
  \bibinfo{pages}{A7} (\bibinfo{year}{2017}).
\newblock \eprint{1508.04114}.

\bibitem{Lykke2017}
\bibinfo{author}{{Lykke}, J.~M.} \emph{et~al.}
\newblock \bibinfo{title}{{The ALMA-PILS survey: First detections of ethylene
  oxide, acetone and propanal toward the low-mass protostar IRAS 16293-2422}}.
\newblock \emph{\bibinfo{journal}{\aap}} \textbf{\bibinfo{volume}{597}},
  \bibinfo{pages}{A53} (\bibinfo{year}{2017}).
\newblock \eprint{1611.07314}.

\bibitem{Calcutt2018b}
\bibinfo{author}{{Calcutt}, H.} \emph{et~al.}
\newblock \bibinfo{title}{{The ALMA-PILS survey: first detection of methyl
  isocyanide (CH$\_3$NC) in a solar-type protostar}}.
\newblock \emph{\bibinfo{journal}{\aap}} \textbf{\bibinfo{volume}{617}},
  \bibinfo{pages}{A95} (\bibinfo{year}{2018}).
%\newblock \emph{\bibinfo{journal}{ArXiv e-prints}}  (\bibinfo{year}{2018}).
\newblock \eprint{1807.02909}.
%}}}1
\end{oldthebibliography}

\clearpage

\begin{addendum}
 \item[Correspondence] Correspondence and requests for materials should be addressed to Jeong-Eun Lee~(email: jeongeun.lee@khu.ac.kr).

 \item  
 ALMA is a partnership of ESO (representing its member states), NSF (USA) and NINS
 (Japan), together with NRC (Canada), NSC and ASIAA (Taiwan), and KASI (Republic
     of Korea), in cooperation with the Republic of Chile. The Joint ALMA Observatory
 is operated by ESO, AUI/NRAO and NAOJ. J.-E.L. is supported by the Basic Science
 Research Program through the National Research Foundation of Korea (grant no. NRF-
     2018R1A2B6003423) and the Korea Astronomy and Space Science Institute under the
 R\&D programme supervised by the Ministry of Science, ICT and Future Planning. G.H.
 is funded by general grant 11473005 awarded by the National Science Foundation of
 China. D.J. is supported by the National Research Council of Canada and by an NSERC
 Discovery Grant. Y.A. acknowledges support from JSPS KAHENHI grant numbers
 16K13782 and 18H05222.

\item[Author Contributions]  
J.-E.L., S.L. and G.B. performed the detailed calculations and line fittings used in the
analysis. J.-E.L. wrote the manuscript. All authors were participants in the discussion of
results, determination of the conclusions and revision of the manuscript.

\item[Competing Interests] The authors declare that they have no competing financial interests.
\end{addendum}

\clearpage
% figures
%{{{1
\begin{figure*}
\includegraphics[height=8cm]{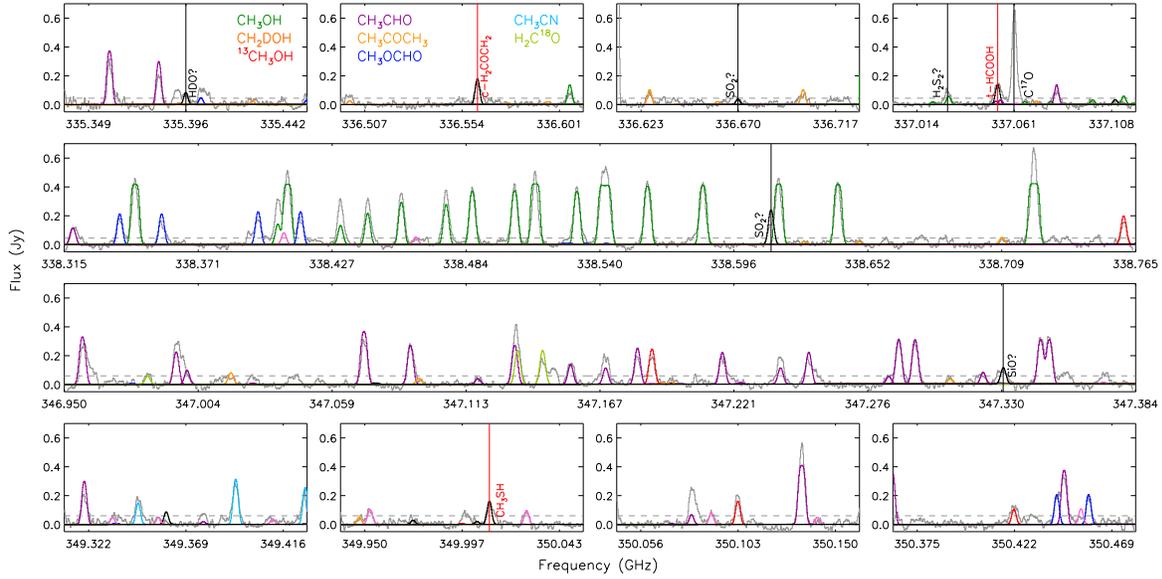}
	\caption{{\bf The COMs detected in the Cycle 5 ALMA observation.} The grey color is for the observed spectra.  The RMS noises of observed spectra are 0.015 and 0.02 Jy for the spectral windows around 330 GHz and the spectral windows around 340 GHz, respectively, and the dashed line indicates the 3 $\sigma$ level.
Different colors indicate the lines of individual species fitted by XCLASS\cite{Moller2017} (see Methods and Supplementary Tables 1 to 3).
 The spectral window in the second row covers the 
  series of CH$_3$OH 7$_k$-6$_k$ transitions. 
 For the line fitting, we adopted  the temperature of 120 K, hydrogen column density of $1.4\times 10^{25}$ cm$^{-2}$, and  the dust properties  
  %$\beta=1$, and $\kappa_{abs}=2.2$ cm$^2$g$^{-1}$ 
from the disk model presented in Methods.
 The identified species and their derived column densities and abundances by this line fitting process are summarized in Supplementary Table 1.
 Ethylene oxide ($c$-H$_2$COCH$_2$), formic acid ($t$-HCOOH), and methyl mercaptan (CH$_3$SH) are tentatively identified; we detect only one transition of each COM.
 The HDO line at 335.395 GHz (upper state energy, $E_u=335$ K) is also marginally detected as marked in the first spectral window, and its estimated column density is 3.8$\times10^{16}$ cm$^{-2}$.
}
\label{fig:fig1}
\end{figure*}

\begin{figure*}
\includegraphics[height=10cm]{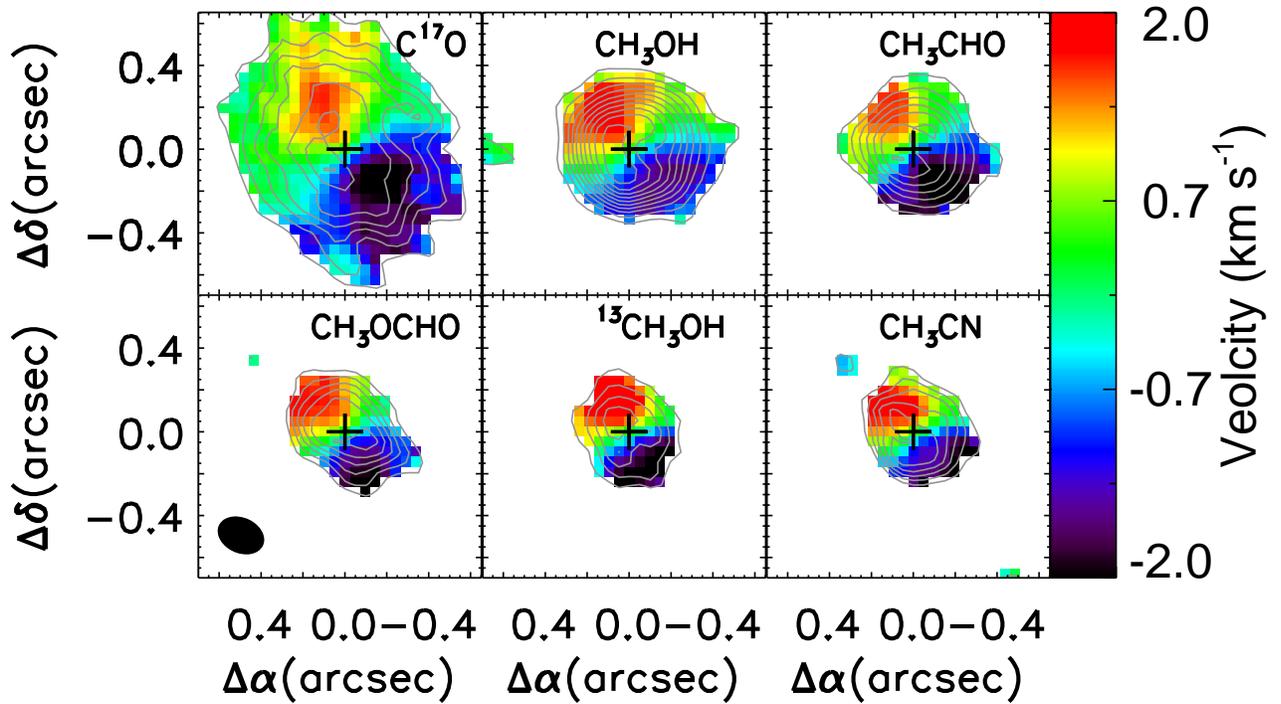}
\caption{{\bf Intensity weighted velocity images (colors) and Integrated intensity maps (contours) of the molecular lines clearly detected in the Cycle 5 ALMA observations.} Multiple lines are detected in each species except for C$^{17}$O, and they are averaged for a high signal-to-noise ratio. The contours are from 5 $\sigma$ in step of 3 $\sigma$ except for CH$_3$OH and CH$_3$CHO. Their contour steps are 5 $\sigma$. The RMS noise levels are 9.9, 3.5, 3.8, 4.5, 6.1, and 7.3 mJy beam$^{-1}$ km s$^{-1}$ for C$^{17}$O 3-2, CH$_3$OH, CH$_3$CHO, CH$_3$OCHO, $^{13}$CH$_3$OH, and CH$_3$CN, respectively. 
%The water snow line identified by Cieza et al. (2016) is marked with the black solid ellipses. 
The cross indicates the position of the continuum peak.
The observed beam is presented in the lower left panel. }
\label{fig:fig2}
\end{figure*}

\begin{figure*}
\includegraphics[height=4.5cm]{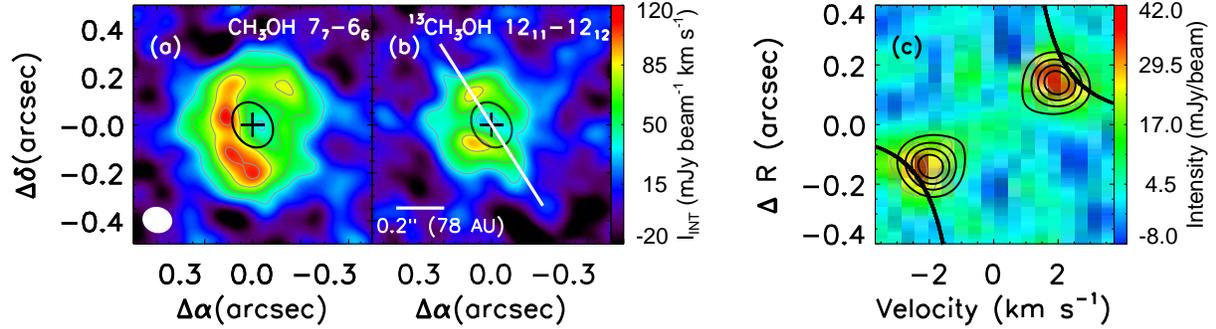}
\caption{{\bf The high-resolution images obtained from the Cycle 4 ALMA observation. a,b,} Integrated intensity (I$_{\rm INT}$) maps of CH$_3$OH 7$_{7}$-6$_{6}$ ({\bf a}) and $^{13}$CH$_3$OH 12$_{11}$-12$_{12}$ ({\bf b}).
A uv-taper was applied to make images with high signal-to-noise ratios and a resolution ($\sim 0.11\arcsec$) enough to resolve clearly the snow line. 
	The lowest contour and subsequent contour step are 5 $\sigma$ and 3 $\sigma$ ($\sigma=$ 7.4 mJy beam$^{-1}$ km s$^{-1}$). The water snow line identified from the analysis of continuum images in Band 6\cite{Cieza2016} is marked with the black solid ellipses.
The synthesized beam is plotted in the left bottom corner. {\bf c,} The position-velocity map along the white line in {\bf b}, which follows the disk semi-major axis.
$\Delta R$ represents the offset from the continuum peak along the white line.
	The thin grey contours are for 5 $\sigma$ and 8 $\sigma$ ($\sigma=$ 4 mJy beam$^{-1}$).
  The thick black solid curves represent the Keplerian motion around the central mass of 1.2 M$_\odot$ with the inclination of 38.3 $^{\circ}$.
  The black contours (20, 40, 60 and 80 \% of the peak intensity) illustrate the synthesized emission from the same simple optically thin disk model for C$^{18}$O 2-1 line\cite{Cieza2016},  except for the beam size and the fact that the emission arises only from 0.1 to 0.2$\arcsec$.}
\label{fig:fig3}
\end{figure*}

\begin{figure*}
\includegraphics[height=5cm]{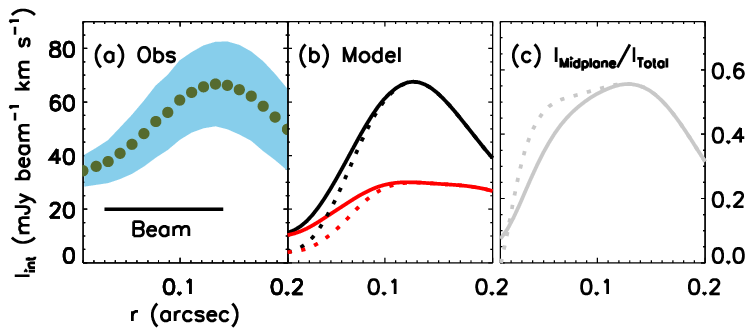}\\
\includegraphics[height=5.5cm]{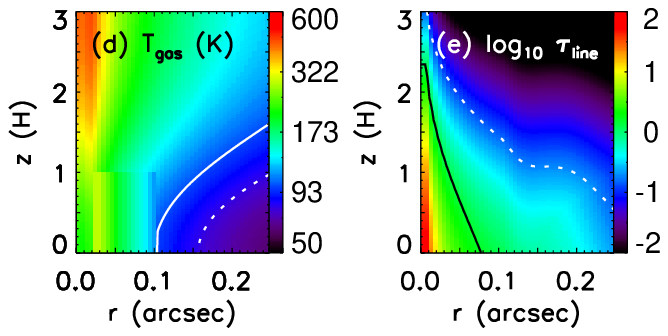}
	\caption{{\bf Models for the $^{13}$CH$_3$OH $12_{11}-12_{12}$ line.} {\bf a,b,} Radial distribution of the observed ({\bf a}) and synthesized ({\bf b}) integrated intensity of $^{13}$CH$_3$OH $12_{11}-12_{12}$. 
Panel {\bf a} shows the azimuthally averaged integrated intensity profile along the de-projected radius r. The sky blue represents the 1 $\sigma$ RMS error. 
In {\bf b}, for each model, the solid lines represent the results from the disk model heated only by irradiation (Model A) while the dotted lines depict the results from the disk model heated by accretion as well as irradiation (Model B). The black lines show the synthesized intensity profile from the entire disk in our models,
while the red lines represent the results from the same models but with  $^{13}$CH$_3$OH only in the disk surface. 
 The difference between black and red lines is equivalent with the emission from the midplane. 
 {\bf c}, The gray lines present the fraction of emission originating from the disk midplane ($I_{\rm Midplane}/I_{\rm Total}$).
{\bf d,e,} The two-dimensional distributions of temperature, $T_{\rm gas}$ ({\bf d}) and optical depth at the line center, $\tau_{\rm line}$ ({\bf e}) for model B. z is the height from the midplane with units of scale height, $H$. The white solid and dashed lines in {\bf d} depict the water and methanol snow lines, respectively. The black solid and white dashed lines in {\bf e} describe the heights with the continuum optical depths of 1 and 0.1, respectively. }
\label{fig:fig4}
\end{figure*}
%}}}1
\clearpage
\noindent{\bf Methods}
%{{{1

\noindent {\bf ALMA observations} 

V883 Ori was observed using the Atacama Large Millimeter/submillimeter Array (ALMA) 
during  Cycle 4  (2016.1.00728.S, PI: Lucas Cieza) on 2017 Sep. 8
and during Cycle 5 (2017.1.01066.T, PI: Jeong-Eun Lee) on 2018 Feb. 21. 
For the Cycle 4 observation, a spectral window was centered at 334.3955GHz 
(HDO 3$_{3,1}$-4$_{2,2}$) with a bandwidth of 93.75 MHz and spectral resolution 
of 488 kHz ($\delta$v$\sim0.4$ km s$^{-1}$). 
Three windows used for continuum, centered at 333.41 GHz, 345.41 GHz, and 347.41 GHz 
with a bandwidth of 2 GHz, are not analyzed in this paper.
The phase center was at 
$(\alpha, \delta)_{J2000}=(05^{h}38^{m}18.10^{s}, -07\degree02'26.0'')$, 
and the total observing time was 15.6 minutes. 
Forty-seven 12-m antennas were used with baselines in the range from 
21 m ($\simeq$24 k$\lambda$) to 10.6 km ($\simeq$11900 k$\lambda$) to provide the 
synthesized beam size of $0.043\arcsec \times 0.025\arcsec$ (PA=$56.4^{\circ}$)
when Briggs weighting (robust = 0.5) is adopted. 
% Cycle 5
For the Cycle 5 observation, ten spectral windows were set to cover many molecular lines 
such as CH$_3$OH, $^{13}$CH$_3$OH, H$^{13}$CO$^+$ \jj43, and C$^{17}$O \jj32. 
Their bandwidths were 468.75 MHz or 117.19 MHz and their spectral resolution was 282 kHz ($\sim$0.2 km s$^{-1}$). The total observing time was 44.53 minutes. 
The phase center was at $(\alpha, \delta)_{J2000}=(05^{h}38^{m}18.10^{s}, -07\degree02'26.0'')$.
Forty-five 12-m antennas were used with baselines in the range from 15 m 
($\simeq$17 k$\lambda$) to 1.4 km ($\simeq$1578 k$\lambda$) to provide 
the synthesized beam size of $0.23\arcsec \times 0.17\arcsec$ (PA=$65.7^{\circ}$) 
when Briggs weighting (robust = 0.0) is adopted.

 We carried out the standard data reduction using CASA 4.7.2 (for Cycle 4) and 5.1.1 (for Cycle 5) \cite{McMullin2007}.
For the Cycle 4 data, the nearby quasar J0541-0541 was used for phase calibration, J0503-1800 for bandpass calibration, and J0423-0120 for flux calibration.
For the Cycle 5 data, J0607-0834 was used for phase calibration while
J0510+1800 was used for bandpass and flux calibration.
% DATA reduction
For the Cycle 4 data, Briggs weighting with a uv-taper of 1000 k$\lambda$ 
was used for the molecular lines to obtain images with a higher signal-to-noise ratio and a resolution ($0.125\arcsec \times 0.108\arcsec$, PA=$75.0^{\circ}$), enough to resolve the water snow line.
 The root mean square (RMS) noise level of the molecular lines is 4 mJy beam$^{-1}$.
For the Cycle 5 data, 
%the visibility data were imaged using the CLEAN algorithm with Briggs weighting (robust = 0.0) and 
the RMS noise level for the molecular lines 
is 7$\sim$8 mJy beam$^{-1}$. We did not use a uv-taper for the Cycle 5 data.

\noindent {\bf Distance to V883 Ori}

V883 is a member of the L1641 cluster\cite{Allen2008}.  A cross-match between the sample of L1641 members\cite{Fang2013} and the Gaia DR2 objects\cite{Gaia2016,Gaia2018} yields 47 stars with parallaxes accurate to at least 5\%, have Gaia $G$-band photometry brighter than 19 mag, are located within 0.5 deg of V883 Ori, and have a proper motion within 1 mas/yr of the average proper motion of the cluster. (The spatial and proper motion cuts are important because the Gaia DR2 astrometry reveals that the L1641 members\cite{Fang2013} include at least two distinct clusters located at different distances.)  This set of stars has a distance of $388$ pc and a standard deviation of 16 pc, roughly consistent with previous a Gaia DR2 analysis\cite{Kounkel2018} .  This standard deviation could be explained by either a real scatter of 12 pc in the distances or by underestimated errors in the parallax measurements.

The Gaia DR2 parallax of V883 Ori yields a distance of $270\pm32$ pc, and a proper motion that is highly discrepant with the L1641 cluster.  If these values are correct, then V883 Ori could not be a member of L1641 or any other known young cluster.  Moreover, the Gaia DR2 catalog flags the astrometric solution of V883 Ori as unreliable, which is not surprising for a variable young star.  We therefore adopt the median distance to the nearby stars in L1641 of 388 pc for V883 Ori.
With this updated distance, the midplane snow line is located at 39 AU rather than 42 AU\cite{Cieza2016} and the mass of the central star is 1.2 M$_\odot$\cite{Cieza2016}.

\noindent {\bf Line Identification and Fitting}

In Figure 1, all lines were extracted with the circular aperture of 0.6\arcsec, within which the signal-to-noise ratio of $^{13}$CH$_3$OH integrated intensity is above 5.
%from each part of disk are combined to reduce the RMS noise.
The line central velocity at each pixel within the aperture, which is affected by the disk rotation, is shifted to the source velocity to reduce the line blending\cite{Yen2016}.
The source velocity is 4.3 km~s$^{-1}$\cite{Cieza2016}.
We identify spectral line transitions (see Supplementary Table 3) and fit the spectra using the eXtended CASA Line Analysis Software Suite (XCLASS)\cite{Moller2017}, which accesses the Cologne Database for Molecular Spectroscopy (CDMS)\cite{Muller2001, Muller2005} and Jet Propulsion Laboratory (JPL)\cite{Pickett1998} molecular databases. 
 We still have several unidentified line transitions. 
In addition, only one transition was detected for each of $c$-H$_2$COCH$_2$, $t$-HCOOH, and CH$_3$SH.

XCLASS takes into account the line opacity and line blending in the assumption of  local thermodynamic equilibrium (LTE). 
The main parameters used to fit lines in XCLASS
are the emission size, the excitation temperature, the column density of the species, the line full width at half maximum (FWHM), and the velocity offset with respect to the systemic velocity of the object. 
The aperture size (0.6\arcsec) was adopted as the emission size.
There is no velocity offset because all lines were shifted to the source velocity before combined.  All lines are assumed to be Gaussian profiles with a FWHM fixed at 2 km s$^{-1}$, the average value of all lines when allowed to vary. 
The excitation temperature of 120 K is adopted from the disk model below; the temperature was averaged within a radius of 0.3\arcsec\ in the model. 
With these parameters, we used MAGIX (Modeling and Analysis Generic Interface 
for eXternal numerical codes)\cite{Moller2013} to
optimize the fit and find the best solution. 

 The observational uncertainty is dominated by the absolute calibration error of 10 \%. The COMs column densities derived from this line fitting are summarized in Supplementary Table 1. The error of the column density is calculated by taking absolute calibration uncertainty and fitting error into consideration.
 The derived column density of $^{13}$CH$_3$OH is larger than that from the excitation diagram (Supplementary Figure 1) by a factor of 7.6. This may be caused by the continuum opacity rather than the optical depths of the $^{13}$CH$_3$OH lines, which are optically thin according to the XCLASS fitting.
The line intensity is scaled by $e^{-\tau^c}$, where $\tau^c$ is the continuum optical depth, and $\tau^c\sim1.4$ in band 7 at 0.1\arcsec. 
 This continuum optical depth effect is taken into account in XCLASS but not in the excitation diagram.

The derived column density of CH$_3$OH suggests a very low $^{12}$C/$^{13}$C ratio of 7.6, which is much lower than the typical ISM value (60)\cite{Langer1993}. The best-fit column density of CH$_3$OH might be underestimated as the modeled lines at the frequency range of 338.42 to 338.48 GHz are weaker than the observed ones. If we increase the column density by a factor of 2, those lines are fitted better. However, this column density does not fit the rest of lines, and it cannot increase the isotopic ratio to the normal value either. This low isotopic ratio between CH$_3$OH and $^{13}$CH$_3$OH may be caused by very high optical depth of CH$_3$OH, which prevents the CH$_3$OH lines from tracing the matter close to the midplane. Alternatively, the 2-D structures of temperature, density, and abundance could be the main reason. The XCLASS fitting of the spectra in Figure 1 is not perfect; the model spectra deviate from the observation for some other lines as well as those methanol lines. It indicates the need for more sophisticated modeling considering the 2-D distributions of temperature, density, and molecular abundances. We postpone such modeling to future work, since our high spatial resolution data cover only a limited number of lines (see Supplementary Table 2).

We also list the COM abundances with respect to CH$_3$OH to compare with the values in the hot corino IRAS 16293B and in comets. In the calculation of the COMs abundances in V883 Ori, we assume that the true column density of CH$_3$OH is the column density of $^{13}$CH$_3$OH multiplied by the elemental abundance of $^{12}$C/$^{13}$C$=60$\cite{Langer1993}. 
 The $^{12}$C/$^{13}$C ratio in CH$_3$OH is expected to be similar to the ratio in CO because CH$_3$OH forms mainly via hydrogenation of CO on grain surfaces \cite{Furuya2011}. Since CO is the dominant carbon carrier, it is reasonable to assume $^{12}$CO/$^{13}$CO=60. 
The $^{12}$CO/$^{13}$CO ratio of 71$\pm$15 measured directly from CO ices in star forming regions\cite{Boogert2002} is also similar to the elemental ratio.
 % The direct measure of the $^{12}$CH$_3$OH/$^{13}$CH$_3$OH ratio is not easy since $^{12}$CH$_3$OH lines tend to be optically thick as found in our fitting. 
 % The $^{12}$C/$^{13}$C ratio measured from the observations of H$_2$CO, which also forms via hydrogenation of CO, is consistent with the elemental ratio\cite{Persson2018}. 
The CH$_3$OH abundance derived from the $^{13}$CH$_3$OH abundance and the $^{12}$C/$^{13}$C ratio of 60 is then $2\times 10^{-7}$, which is in reasonable agreement with the recent study ($4\times10^{-7}$)\cite{vantHoff2018}.

 As listed in Supplementary Table 1, the COMs abundances relative to CH$_3$OH are higher than those in the hot corino IRAS 16293 B and similar to those in comets, except for CH$_3$CN.  This pattern is indicative of chemical evolution in the disk that continues after the ISM material is incorporated to the disk. During the quiescent phase, for example, methyl formate (CH$_3$OCHO) can form via grain-surface reactions of HCO and CH$_3$O in the disk midplane\cite{walsh2014}. However, it is difficult to compare directly our observational results with the chemical models because there are still many uncertainties in the grain surface reactions, such as the branching ratios of radical-radical reactions.

\noindent {\bf Models}

In order to investigate the 2-D (r and z) distribution of CH$_3$OH from our high spatial resolution data, we construct a simple irradiated disk model to calculate the radiative transfer 
of the $^{13}$CH$_3$OH 12$_{11}$-12$_{12}$ line in the assumption of LTE.
%local thermal equilibrium. 
The gas is assumed to be well coupled with dust grains 
in the region where the $^{13}$CH$_3$OH line forms, and thus,  the gas temperature is the same as the dust temperature. We assume that the gas to dust mass ratio is constant as 100 over the entire disk. 
The vertical density profile in the disk is given as
\begin{equation}
 n_{\rm H_2}(r,z) = \frac{N_{\rm H_2}(r)}{H\sqrt{2\pi}}\exp(-\frac{z^2}{2H^2}),
\end{equation}
 with the H$_2$ column density, $N_{\rm H_2}$ and the scale height, $H$.
 For the vertical temperature distribution, we follow the vertical gradient 
prescription of an irradiated disk model\cite{Dartois2003,Andrews2011}:
\begin{eqnarray}
T(r,z) &=&  T_{\rm atm} (r) + (T_{\rm mid} (r) - T_{\rm atm} (r))
        \cos ^4\left(\frac{\pi z}{8 H}\right) \qquad if\, z< 4 H \nonumber\\
        &=&  T_{\rm atm} (r) \qquad  if \,  z \geq 4 H,
\end{eqnarray}
 with the temperature at the atmosphere, $T_{\rm atm}$ and that at the midplane, $T_{\rm mid}$. For simplicity, we assume that $T_{\rm atm}/T_{\rm mid}$ is 2 
 based on the dust continuum radiative transfer model for V883 Ori \cite{Cieza2018}.

To find $N_{\rm H_2}(r)$ and $T_{\rm mid}(r)$, we solve a dust continuum radiative transfer for the 
 high resolution ALMA continuum image at Band 6:
\begin{equation}
 I_{{\nu_0}(r)} = \int_{-\infty}^{+\infty} B_{\nu_0}(T(r,z)) \exp(-\tau^c_{\nu_0}(r,z))\kappa^c_{\nu_0}(r,z) dz,
 \label{eq:rt}
\end{equation}
with 
\begin{eqnarray}
 \kappa^c_{\nu_0}(r,z) &=&  \kappa_{\rm abs}\,m_{\rm H_2} n_{\rm H_2}(r,z)/g2d \\ \nonumber
 \tau^c_{\nu_0}(r,z) &=& \int_{-\infty}^z   \kappa^c_{\nu_0}(r,z')\,dz'.
\end{eqnarray}
where $\kappa_{\rm abs}$ is the dust absorption opacity at 1.3 mm ($2.2~{\rm cm^2 g^{-1}}$)\cite{Cieza2018},
$m_{\rm H_2}$ is the molecular hydrogen mass, 
  and $g2d$ is the gas to dust mass ratio.

 A previous study\cite{Cieza2016} shows that V883 Ori has a very optically thick inner disk and a optically thin outer disk. 
The boundary, that is located at 0.1\arcsec, is interpreted as water snow line.
  In the inner disk, the dust temperature is relatively well constrained while the dust temperature and the column density are degenerate 
    in the optically thin outer disk. Therefore, we analyze the inner and outer disks separately.
%and for the outer disk, we calculate the column density in a simple assumption of temperature.
%The water snow line is located at 0.1\arcsec in the midplane,
 %{\bf and inward the snow line, the continuum optical depth increases sharply\cite{Cieza2016}. }
% Therefore, we derive the temperature and {\bf column density} profiles separately inward and outward the water snow line at 0.1\arcsec. 
  At $r>$ 0.1\arcsec, we assume that $T_{\rm mid}(r)$ follows the profile of 
 $T_{\rm mid} (0.1\arcsec) \times \sqrt{0.1\arcsec/r}$, as used in the irradiated disk\cite{DAlessio1998}. We adopt $T_{\rm mid} (0.1\arcsec) = 100$ K (ref. \cite{Cieza2016}) based on the temperature estimated in the optically thick region. Given temperature distribution, $N_{\rm H_2}(r)$ can be derived by solving above equations.
 The derived column density at $0.1\arcsec$ is $\sim 1.3 \times 10^{25}$~cm$^{-2}$ and the density at the midplane is $\sim 10^{11}$~cm$^{-3}$ when $H/r$ is 0.1.
  At this density, the freeze-out timescale of COMs is shorter than a year if the temperature is lower than the sublimation temperature\cite{Lee2004}.
To derive the temperature profile at the inner disk ($r \le 0.1\arcsec$), $N_{\rm H_2}(r)$ is set to be 
$N_{\rm H_2}(0.1\arcsec) \times (r/0.1\arcsec)^{-\gamma}$. 
Here, we adopt $\gamma = 1.5$, a typical power index of the surface density profile in the disk\cite{Schoonenberg2017}. 
 The temperature distribution in this disk model (Model A, hereinafter) is shown in Supplementary Figure 4.

%The derived temperature profile ($T_{\rm dust}$) and the corresponding continuum optical depth at Band 6 are presented in the black solid line in Supplementary Figure 4 (leftmost panel). 
%The black dotted line shows the extension of the temperature profile at $r>0.1\arcsec$ inward the snow line. At $r<0.1\arcsec$, the temperature derived from the observation (solid line) is higher than the temperature of the irradiated disk (dotted line), indicative of an additional heating process (e.g., accretion).

For the synthetic $^{13}$CH$_3$OH line emission, the continuum absorption coefficient is considered as a function of frequency,
$\kappa^c_\nu (r,z)$ = $\kappa^c_{\nu_0} (r,z) \times \left(\nu/\nu_0\right)^{\beta}$ with the spectral 
index of $\beta=1$ in the entire disk. The total absorption coefficient is given by 
\begin{eqnarray}
\kappa_\nu (r,z) &=& \kappa^c_\nu (r,z) + \kappa^l_\nu(r,z), \\ \nonumber
\kappa^l_\nu(r,z)&=& \frac{c^3}{8 \pi \nu_{ul}^3}
    \frac{A_{ul}}{1.064 \Delta V}
    \frac{n_{\rm H_2}(r,z) X_{\rm mol}}{Q(T)} 
    \left[\chi_l \frac{g_u}{g_l} - \chi_u \right] 
    \exp \left[-\frac{1}{2} \left(\frac{\Delta v}{\sigma}\right)^2\right],
\end{eqnarray}
where $c$ is the speed of light, $\nu_{ul}$ and $A_{ul}$ are the 
frequency and the Einstein coefficient of the transition, respectively,
 $\chi_u$ ($\chi_l$) and $g_u$ ($g_l$) is the level population and statistical weight in the upper 
(lower) energy state, respectively. 
$\Delta V$ (= $\sigma \times 2 \sqrt{2 \ln 2}$; in cm~s$^{-1}$) is the full width at half maximum of the line profile,  
$\Delta v$ is the velocity shift from the source velocity, 
$Q(T)$ is the partition function, 
and $X_{\rm mol}$ is the molecular abundance of $^{13}$CH$_3$OH. 
We adopted $\Delta V=2$ km s$^{-1}$, which is the average value over all lines, and $X_{\rm mol}=4\times 10^{-9}$, which reproduces the observed peak intensity. When $T_{\rm gas}$ is lower than the sublimation temperature, 
the molecule is assumed to be frozen onto grain surfaces, depleted from the gas.
The sublimation temperature of $^{13}$CH$_3$OH, which fits the radial intensity profile, was 80 K, as marked with the white dashed line in Figure 4(d); above the dashed line, $X_{\rm mol}=4\times 10^{-9}$, but below the line, $X_{\rm mol}=0$. This lower sublimation temperature of CH$_3$OH than water sublimation temperature is consistent with the lower binding energy  (5000 K) of CH$_3$OH than that of water (5600 K)\cite{Wakelam2017}.
The molecular data are adopted from CDMS database\cite{Muller2001}.

Finally, the line emission is synthesized using the equation (3), but with the total absorption coefficient, $\kappa_\nu (r,z)$, instead of $\kappa^c_{\nu_0}$. 
The calculated intensity profile is smoothed to have the same resolution (0.11\arcsec) of the observed image.  
To check whether the molecular line can trace down to the midplane, 
 we also calculated a model in which the CH$_3$OH abundance is artificially set to be zero in the midplane ($|z| \le H$) (red lines in Figure 4); the difference between the black and red lines can be considered the emission from the disk midplane. The fraction of emission from the disk midplane is presented by the gray lines in Figure 4.

 The spectral index derived from the high resolution Band 7 and Band 6 continuum images of V883 Ori is smaller than 2, 
 reaching down to 1, within $\sim 0.1\arcsec$ (Cassasus et al. in prep.). In the submillimeter regime, the free-free emission is negligible,
 and thus, most flux is likely from thermal emission, whose spectral index must be greater than 2.
 Note that we can see a deeper region in Band 6 than in Band 7. Thus, the flux ratio of Band 7 to Band 6 smaller than expected from 
 the isothermal case suggests that the deeper region is hotter. 
Therefore, we model the disk heated by accretion (at $r<0.1\arcsec$) as well as irradiation (Model B). 

 In Model B, to combine the two heating processes, irradiation and accretion, in a simple way, we assume that $N_{\rm H_2}(r)$ 
  and $T_{\rm mid} (r)$ are the same as those of Model A, except for the temperature distribution within the snow line. 
  Within the snowline ($r < 0.1\arcsec$), we assume that the midplane is vertically isothermal at $z\le H$ (accretion heating), while the temperature gradient is given by Equation (2) at $z > H$ (irradiative heating).
 In Equation (2), which describes the temperature profile resulted by irradiation, we assume $T_{\rm mid} (r)= T_{\rm mid} (0.1\arcsec) \times \sqrt{0.1\arcsec /r}$ as used for 
    $r> 0.1\arcsec$. 
In Model A at a given radius, one temperature was found to fit the continuum intensity. Therefore,
if at a given radius, the $T_{\rm mid}(r)$ calculated by the above equation is lower than $T_{\rm mid}(r)$ in Model A, then a higher 
    midplane temperature (the blue line in the upper right panel in Supplementary Figure 4) is found to fit the observed intensity. 
    The derived temperature distribution of Model B is presented in  Figure 4(d).
%}}}1
\clearpage

\begin{addendum}
 \item[Data availability]
%This paper makes use of the following ALMA data: ADS/JAO.ALMA\#2016.1.00728.S, ADS/JAO.ALMA\#2017.1.01066.T. 
This paper makes use of the ALMA data,  which could be downloaded from the ALMA archive (https://almascience.nao.ac.jp/aq/) with project codes 2016.1.00728.S and 2017.1.01066.T.
The data that support the plots within this paper and other findings
of this study are available from the corresponding author upon reasonable request.

\end{addendum}

%\bibliography{v883ori}

%}}}1

%%
%% TABLES
%%
%% If there are any tables, put them here.
%%

\clearpage

 \scriptsize
\tiny
\begin{ThreePartTable}
\begin{TableNotes}
\item[a] \label{a}{\bf The uncertainty of column density corresponds to the 1 $\sigma$ error.}
\item[b] \label{b}X($w.r.t$ CH$_3$OH) was calculated using the derived column density of $^{13}$CH$_3$OH and the $^{12}$C/$^{13}$C ratio of 60.
%\item[c] \label{c}X($w.r.t$ CH$_3$OH) from the refs\cite{Jorgensen2016,Lykke2017,Calcutt2018b} 
\item[c] \label{c}X($w.r.t$ CH$_3$OH) from the refs (8,29,30)%\cite{Jorgensen2016,Lykke2017,Calcutt2018b} 
%\item[d] \label{e}X($w.r.t$ CH$_3$OH) from the ref\cite{LeRoy2015}
\item[d] \label{e}X($w.r.t$ CH$_3$OH) from the ref (4) %\cite{LeRoy2015}
\item[e] \label{e}Only one transition was detected for these COMs. Therefore, these three COMs are tentatively identified. 
\end{TableNotes}
%\begin{tabularx}{1.\textwidth}{ccccccc}
\begin{tabularx}{1.\textwidth}{c *{7}{Y}}
\multicolumn{7}{c}{\bf Supplementary Table 1: Identified COMs}
\label{tb:sp_table1}
\\\hline
{Species} &
{Formula} &
{Column density\tnote{a}} &
{X({\it w.r.t.} H$_2$)}&
{X({\it w.r.t.} CH$_3$OH)\tnote{b}}&
{IRAS16293B\tnote{c}}&
{Comets\tnote{d}}&\\
\hline\hline
Methanol & CH$_{3}$OH &  4.00$_{-0.44}^{+0.96}$ $\times$ 10$^{17}$ & 2.86 $\times$ 10$^{-8}$ &  -- & -- & --\\
         & CH$_{2}$DOH & 6.21$_{-0.74}^{+0.69}$ $\times$ 10$^{16}$ & 4.43 $\times$ 10$^{-9}$ & 1.97$\times$ 10$^{-2}$ &-- &-- \\
         & $^{13}$CH$_{3}$OH & 5.26$_{-0.55}^{+0.53}$ $\times$ 10$^{16}$ & 3.76 $\times$ 10$^{-9}$ & 1.67 $\times$ 10$^{-2}$ &--& --\\
Acetaldehyde & CH$_{3}$CHO & 6.40$_{-0.69}^{+0.64}$  $\times$ 10$^{16}$ & 4.57 $\times$ 10$^{-9}$ & 2.03$\times$ 10$^{-2}$&3.50$\times$ 10$^{-3}$ &  1.0 -- 4.4 $\times$ 10$^{-2}$ \\
Methyl Formate & CH$_{3}$OCHO & 2.37$_{-0.24}^{+0.24}$ $\times$ 10$^{17}$ & 1.69 $\times$ 10$^{-8}$ &  7.50$\times$ 10$^{-2}$& 2.00$\times$ 10$^{-2}$ & 1.3 -- 4.2 $\times$ 10$^{-2}$ \\
Acetone & CH$_{3}$COCH$_{3}$ & 4.32 $_{-0.46}^{+0.45}$  $\times$ 10$^{16}$ & 3.08 $\times$ 10$^{-9}$ & 1.37$\times$ 10$^{-2}$ &8.50$\times$ 10$^{-4}$  &-- \\
Acetonitrile & CH$_{3}$CN & 2.45$_{-0.25}^{+0.26}$ $\times$ 10$^{15}$ & 1.75 $\times$ 10$^{-10}$ & 7.76$\times$ 10$^{-4}$&  2.00$\times$ 10$^{-3}$& 5.0 -- 53.0$\times$ 10$^{-3}$ \\
Ethylene oxide & $c$-H$_2$COCH$_2$\tnote{e} & 3.73$_{-0.41}^{+0.38}$ $\times$ 10$^{16}$ & 2.67 $\times$ 10$^{-9}$ & 1.18$\times$ 10$^{-2}$ &-- &-- \\
Formic acid & $t$-HCOOH\tnote{e} &  1.00$_{-0.10}^{+0.11}$ $\times$ 10$^{18}$ & 7.16 $\times$ 10$^{-8}$ & 3.18$\times$ 10$^{-1}$ &-- &-- \\
Methyl mercaptan & CH$_3$SH\tnote{e} & 8.83$_{-1.00}^{+1.10}$ $\times$ 10$^{16}$ & 6.31 $\times$ 10$^{-9}$ & 2.80$\times$ 10$^{-2}$ &-- &-- \\
%Water & HDO\tnote{d} & 3.77$_{-0.43}^{+0.64}$ $\times$ 10$^{16}$ & 2.69 $\times$ 10$^{-9}$ & 1.19$\times$ 10$^{-2}$ &-- &-- \\
\hline
 \insertTableNotes
\end{tabularx}
 \end{ThreePartTable}
% \normal
\clearpage

 \tiny
\scriptsize
\begin{longtable}{ccccccc}
%\begin{tabularx}{\textwidth}{c *{7}{Y}} 
%\begin{table}{ccccccc}
\multicolumn{7}{c}{\bf Supplementary Table 2: Cycle 4 Line Identification}
\label{tb:sp_table2_cycle4}
\\\hline
{Line No.} &
{Formula} &
{Name} &
{Frequency [GHz]}&
{Transition}&
{Einstein-A [log$_{10}$A]}&
{ E$_{\rm u}$ [K]}
\\\hline\hline
%\multicolumn{7}{c}{Cycle 4}\\
%\hline
1 &  CH$_{3}$CHO vt = 1 & Acetaldehyde &    334.98085310 (3.48E-5) &    18(1, 18) - 17(1, 17), E &    -2.87593 &    359.89377\\
2 & HDCO &    Formaldehyde &    335.09673940 (8.96E-5) &    5( 1, 4)- 4( 1, 3)   &  -2.98087   &  56.24826\\
3 & CH$_{3}$OH vt = 0 &    Methanol   &  335.13357000 (1.3E-5) &     2(2)- - 3(1)- &   -4.57254 &    44.67266\\
4 & CH$_{3}$CHO v = 0 &   Acetaldehyde &    335.31810910 (2.83E-5) &      18(0, 18) - 17(0, 17), E   &  -2.87112   &  154.92723\\
5 & CH$_{3}$CHO v = 0 &   Acetaldehyde &    335.35872250 (2.83E-5) &      18(0, 18) - 17(0, 17), E &    -2.87128 &    154.85292\\
6 & CH$_{3}$CHO vt = 1 &    Acetaldehyde   &  335.38246150 (3.91E-5) &    18(0, 18) - 17(0, 17), E   &  -2.86462 &    361.48406\\
7 &$^{13}$CH$_{3}$OH vt = 0 &   Methanol   &  335.56020700 (4.0E-5) &     12( 1, 11)- 12( 0, 12) - + &    -3.39358 &    192.65405\\
8 &CH$_{3}$OH vt = 0 &    Methanol &    335.58201700 (5.0E-6) &   7(1)+ - 6(1)+  &  -3.78844   &  78.97183\\
9 & H$_{2}$C$^{18}$O   &  Formaldehyde   &  335.81602540 (2.40E-4) &    5( 1, 5)- 4( 1, 4) &    -2.97975   &  60.23580\\
\hline
\end{longtable}
%\end{tabularx}

\clearpage

\tiny
\begin{longtable}{>{\centering}p{0.08\textwidth}cccccc}
\multicolumn{7}{c}{\bf Supplementary Table 3: Cycle 5 Line Identification}
\label{tb:sp_table2_cycle5}
\\\hline
{Line No.} &
{Formula} &
{Name} &
{Frequency [GHz]}&
{Transition}&
{Einstein-A [log$_{10}$A]}&
{ E$_{\rm u}$ [K]} 
\\\hline\hline
\multicolumn{7}{c}{spw0}\\
\hline
1 &CH$_{3}$CHO v=0 &Acetaldehyde &335.35872250 (2.83E-5) &18(0, 18) - 17(0, 17), E &-2.87128 &154.85292 \\
2 &CH$_{3}$CHO vt=1	& Acetaldehyde &	335.38246150 (3.91E-5)& 	18(0, 18) - 17(0, 17), E &	    -2.86462 & 361.48406 \\
3 & HDO ? &	Water &	 	335.39550000 (2.6E-5) & 3( 3, 1)- 4( 2, 2) &	-4.58367 &	335.26718 \\
\hline
\multicolumn{7}{c}{spw1}\\
\hline
4 & c-H$_{2}$COCH$_{2}$  &	Ethylene Oxide &	 	336.56139300 (0.00015) & 9( 4, 5)- 8( 5, 4) &	-3.52971 &	90.31306 \\
5 & CH$_{3}$OH vt=2 & Methanol             &	      336.60588900 (1.3E-5) &	 	7(1)+ - 6(1)+ &	-3.78629	&	747.41346 \\
\hline
\multicolumn{7}{c}{spw2}\\
\hline
6 & (CH$_{3}$)$_{2}$CO v=0 &	Acetone &	336.62703300  (2.76E-5) & 17(16, 1)-16(15, 1) EE & -2.80789  &	142.45589 \\
7 &SO$_{2}$ v = 0 ? &		Sulfur dioxide &		336.66957740 (3.9E-6) &	16( 7, 9)-17( 6,12) &		-4.23374 &		245.11422 	\\
8 &(CH$_{3}$)$_{2}$CO v=0 &		Acetone &		336.70098190 (3.14E-5) &	 	17(16, 2)-16(15, 2) EE	 &	-2.80766 &		142.34420\\
\hline
\multicolumn{7}{c}{spw3}\\
\hline
9 & H$_{2}$S$_{2}$ ?&		Hydrogen disulfide &		337.02909220 (6.0E-6) &	25(3,22) - 26(2,24) &		-4.25089 &		277.68170\\
10&CH$_{3}$OH vt=2 &	Methanol	& 337.02957300 (1.8E-5) &	7(2) - 6(2) & -3.80855	& 941.38794 \\
11 & $t$-HCOOH & Formic Acid & 337.05330470 (4.07E-5) & 11( 3, 9)-11( 2,10) & -4.61210 & 58.20290 \\
12 & C$^{17}$O &		Carbon Monoxide	 &	 	337.06112980 (1.0E-5) &		J=3-2 &		-5.63440 &		32.35323\\
13 &CH$_{3}$CHO vt = 2 &		Acetaldehyde &		337.08157220 (0.0001427) &		18(1, 18) - 17(1, 17), E &		-2.88192	 &	526.13974  \\
\hline
\multicolumn{7}{c}{spw4}\\
\hline
14 & CH$_{3}$CHO vt = 2	  &		Acetaldehyde &		338.31785810 (0.000161) &		18(5, 14) - 17(5, 13), E	 &	-2.86939  &		605.15649 \\
15 & CH$_{3}$CHO vt = 2	  &		Acetaldehyde	 &	338.31906720 (0.000161) &	 	18(5, 13) - 17(5, 12), E	 &	-2.86939	 &	605.15655\\
16 &		 $^{34}$SO$_{2}$  v=0	? &	Sulfur Dioxide &		338.32036020 (4.4E-6) &	13(2,12)-12(1,11) &		-3.64436 &		92.58313 \\
17 &		CH$_{3}$OCHO v=0 &		Methyl Formate	 &	 	338.33818400 (0.0002) &		27( 8,19)-26( 8,18) E	 &	-3.26914 &		267.18358	 \\
18 &		CH$_{3}$OH vt = 0 &		Methanol &		338.34458800 (5.0E-6) &	 	7(-1,7) - 6(-1,6) &		-3.77807 &	70.55083  \\
19 &		CH$_{3}$OCHO v=0 &		Methyl Formate	 &		338.35579200 (0.0001) &		27( 8,19)-26( 8,18) A &		-3.26882 &		267.18572 \\	
20 &		CH$_{3}$OCHO v=0 &		Methyl Formate	 &	 	338.39631800 (0.0001) &		27( 7,21)-26( 7,20) E &		-3.25943 &		257.74503 \\
21 &		CH$_{3}$OH vt = 0 &		Methanol &		338.40461000 (5.0E-6) &		 	7(6) - 6(6), E1 &		-4.34500	 &	243.79221 \\
22 &		CH$_{3}$OH vt = 0 &		Methanol &		338.40869800 (5.0E-6) &		7(0)+ - 6(0)+ &		-3.76895 &		64.98149 \\
23 &		CH$_{3}$OCHO v=0 &		Methyl Formate	 &	 	338.41411600 (0.0001) &		27( 7,21)-26( 7,20) A	 &	-3.25931 &		257.74703 \\
24 &		CH$_{3}$OH vt = 0 &		Methanol	 &	338.43097500 (5.0E-6) &		 	7(-6) - 6(-6), E2	 &	-4.34266	 &	253.94843 \\
25 &		CH$_{3}$OH vt = 0 &		Methanol &		338.44236700 (5.0E-6) &		7(6)+ - 6(6)+ &		-4.34360	 &	258.69841 \\	
26 &		CH$_{3}$OH vt = 0 &		Methanol &		338.45653600 (5.0E-6) &		7(-5) - 6(-5), E2	 &	-4.07836 &		188.99997 \\	
27 &		CH$_{3}$OH vt = 0 &		Methanol &		338.47522600 (5.0E-6) &	 	7(5) - 6(5), E1 &		-4.07807 &		201.06077	 \\
28 &		CH$_{3}$OH vt = 0 &		Methanol &		338.48632200 (5.0E-6) &		 	7(5)+ - 6(5)+	 &	-4.07635 &		202.88569	 \\
29 &		CH$_{3}$OH vt = 0 &		Methanol &		338.50406500 (5.0E-6) &		 	7(-4) - 6(-4), E2	 &	-3.94036	 &	152.89447 \\	
30 &		CH$_{3}$OH vt = 0 &		Methanol &		338.51263200 (5.0E-6) &		7(4) - 6(4)	 &	-3.93970	 &	145.33406 \\
31 &		CH$_{3}$OH vt = 0 &		Methanol &		338.51264400 (5.0E-6) &		7(4)+ - 6(4)+ &		-3.93970	 &	145.33406 \\
32 &		CH$_{3}$OH vt = 0 &		Methanol &		338.51285300 (5.0E-6) &		 	7(2)- - 6(2)- &		-3.80281	 &	102.70283 \\	
33 &		CH$_{3}$OH vt = 0 &		Methanol &		338.53025700 (5.0E-6) &		 	7(4) - 6(4), E1 &		-3.93781 &		160.99178 \\	
34 &		CH$_{3}$OH vt = 0 &		Methanol &		338.54082600 (5.0E-6) &	 	7(3)+ - 6(3)+ &		-3.85727 &		114.79429	 \\
35 &		CH$_{3}$OH vt = 0 &		Methanol &		338.54315200 (5.0E-6) &	 	7(3)- - 6(3)-	 &	-3.85727 &		114.79441	 \\
36 &		CH$_{3}$OH vt = 0 &		Methanol &		338.55996300 (5.0E-6) &		 	7(-3) - 6(-3), E2 &		-3.85416 &		127.70688	 \\
37 &		CH$_{3}$OH vt = 0 &		Methanol &		338.58321600 (5.0E-6) &		7(3) - 6(3), E1	 &	-3.85584	 &	112.71009	 \\
38 &		SO$_{2}$ v = 0 ? &		Sulfur dioxide &		338.61180780 (3.5E-6) &	 	20( 1,19)-19( 2,18)	 &	-3.54241 &		198.87750 \\	
39 &		CH$_{3}$OH vt = 0	 &	Methanol &		338.61493600 (5.0E-6) &		 	7(1) - 6(1), E1	 &	-3.76659	 &	86.05234	 \\
40 &		CH$_{3}$OH vt = 0	 &	Methanol &		338.63980200 (5.0E-6) &	 	7(2)+ - 6(2)+ &		-3.80233	 &	102.71756 \\	
41 &		CH$_{3}$OH vt = 0	 &	Methanol &		338.72169300 (5.0E-6) &		 	7(2) - 6(2), E1	 &	-3.80941 &		87.25885	 \\
42 &		CH$_{3}$OH vt = 0	 &	Methanol &		338.72289800 (5.0E-6) &		 	7(-2) - 6(-2), E2 &		-3.80422 &		90.91342 \\
43 & $^{13}$CH$_{3}$OH vt = 0	 &	Methanol &		338.75994800 (5.0E-5) &	 	13( 0, 13)- 12( 1, 12) + +	 &	-3.66184 &		205.94501 \\
\hline
\multicolumn{7}{c}{spw5}\\
\hline
44  & CH$_{3}$CHO v = 0	 &	Acetaldehyde &		346.95755590 (2.84E-5) &		 	18(7, 12) - 17(7, 11), E &		-2.89625	 &	268.60623 \\
    & CH$_{3}$CHO v = 0	 &	Acetaldehyde &		346.95755770 (2.84E-5) &		 	18(7, 11) - 17(7, 10), E &		-2.89625 &		268.60624 \\
45 & H$_{2}$C$^{18}$O &		Formaldehyde &		346.98406710 (0.0002806) &		 	5( 4, 2)- 4( 4, 1) &		-3.36307 &		239.79906\\
46 & H$_{2}$C$^{18}$O &		Formaldehyde &		346.98409360 (0.0002806) &		 	5( 4, 1)- 4( 4, 0) &		-3.36307 &		239.79906 \\	
47 & CH$_{3}$CHO v = 0	 &	Acetaldehyde &		346.99553240 (2.82E-5) &		 	18(7, 12) - 17(7, 11), E	 &	-2.89625 &		268.57209 \\
48 & CH$_{3}$CHO vt = 2	 &	Acetaldehyde &		346.99991340 (0.0001791) &		 	18(7, 11) - 17(7, 10), E	 &	-2.89745 &		646.35557 \\	
   & CH$_{3}$CHO vt = 2	 &	Acetaldehyde	 &	346.99994090 (0.0001791) &	 	18(7, 12) - 17(7, 11), E	 &	-2.89745 &		646.35557 \\	
49 & CH$_{3}$CHO v = 0	 &	Acetaldehyde &		347.07154710 (2.58E-5) &		 	18(6, 13) - 17(6, 12), E	 &	-2.87569	 &	239.39974 \\	
   & CH$_{3}$CHO v = 0	 &	Acetaldehyde &		347.07168440 (2.58E-5) &		 	18(6, 12) - 17(6, 11), E &		-2.87569 &		239.39975 \\	
50 & CH$_{3}$CHO v = 0	 &	Acetaldehyde &		347.09040150 (2.57E-5) &		 	18(6, 12) - 17(6, 11), E	 &	-2.87577 &		239.39762 \\
51 & CH$_{3}$CHO v = 0	 &	Acetaldehyde &		347.13268590 (2.58E-5) &	 	18(6, 13) - 17(6, 12), E	 &	-2.87563	 &	239.32124 \\
52 & H$_{2}$C$^{18}$O & Formaldehyde &		347.13389950 (0.0002025) &		 	5( 3, 3)- 4( 3, 2)	 &	-3.11271 &		156.76944  \\
53 & H$_{2}$C$^{18}$O & Formaldehyde &		347.14404620 (0.0002025) &		5( 3, 2)- 4( 3, 1) &		-3.11259 &		156.77007 \\	
54 & CH$_{3}$CHO vt = 2 &		Acetaldehyde &		347.15512480 (0.0001461) &	 	18(4, 14) - 17(4, 13), E	 &	-2.84730 &		573.92991 \\	
55 & CH$_{3}$CHO v = 0	 &	Acetaldehyde &		347.16951960 (3.96E-5) &	 	19(0, 19) - 18(1, 18), E	 &	-3.66327 &		171.81487	 \\
56 & CH$_{3}$CHO vt = 1	 &	Acetaldehyde &		347.18241320 (2.99E-5) &		 	18(4, 14) - 17(4, 13), E	 &	-2.84531	 &	400.37805 \\	
57 & $^{13}$CH$_{3}$OH vt = 0	 &	Methanol &		347.18828300 (6.4E-5) &	 	14( 1, 13)- 14( 0, 14) - + &		-3.36086 &		254.25185 \\
58 & CH$_{3}$CHO vt = 1	 &	Acetaldehyde &		347.21679780 (2.88E-5) &		 	18(5, 13) - 17(5, 12), E &		-2.85923 &		420.44041 \\
59 & CH$_{3}$CHO v = 0	 &	Acetaldehyde &		347.23739280 (7.6E-6) &		 	8(2, 6) - 7(0, 7), E &		-5.82082	 &	42.53799	 \\
60 & CH$_{3}$CHO vt = 1	 &	Acetaldehyde &		347.25182200 (3.08E-5) &		 	18(5, 14) - 17(5, 13), E	 &	-2.85890	 &	419.67263 \\
61 & CH$_{3}$CHO vt = 1	 &	Acetaldehyde &		347.28421220 (7.31E-5) &		 	18(10, 8) - 17(10, 7), E &		-2.98719 &		586.88461 \\	
62 & CH$_{3}$CHO v = 0	 &	Acetaldehyde &		347.28826400 (2.48E-5) &	 	18(5, 14) - 17(5, 13), E &		-2.85868	 &	214.69755 \\	
63 & CH$_{3}$CHO v = 0	 &	Acetaldehyde &		347.29487350 (2.48E-5) &		 	18(5, 13) - 17(5, 12), E	 &	-2.85858	 &	214.69830 \\	
64 & CH$_{3}$CHO vt = 1	 &	Acetaldehyde &		347.32243660 (5.17E-5) &		 	18(9, 9) - 17(9, 8), E	 &	-2.95129	 &	543.81724 \\	
65 & SiO v = 0	 &	Silicon Monoxide &		347.33063100 (0.0003475) &		 	8-7	 &	-2.65578 &		75.01697	 \\
66 & CH$_{3}$CHO v = 0	 &	Acetaldehyde &		347.34571040 (2.48E-5) &		 	18(5, 13) - 17(5, 12), E	 &	-2.85860 &		214.64074 \\
67 & CH$_{3}$CHO v = 0	 &	Acetaldehyde &		347.34927830 (2.48E-5) &	 	18(5, 14) - 17(5, 13), E &		-2.85854 &		214.61141	 \\
\hline
\multicolumn{7}{c}{spw6}\\
\hline
68 & CH$_{3}$CHO vt = 1	 &	Acetaldehyde &		349.32035150 (3.33E-5) &		 	18(1, 17) - 17(1, 16), E	 &		-2.81548		 &	369.25641 \\
69 & CH$_{3}$CN v = 0	 &	Methyl Cyanide &		349.34634280 (2.0E-7) &		19(4) - 18(4)		 &	-2.61106	 &		281.98778 \\
70 & CH$_{3}$CN v = 0	 &	Methyl Cyanide &		349.39329710 (2.0E-7) &		 	19(3) - 18(3)	 &		-2.60218		 &	232.01121 \\
\hline
\multicolumn{7}{c}{spw7}\\
\hline
71 & CH$_{2}$DOH	 &	Methanol &		349.95168460 (7.1E-6) &		 	7(4,4) - 7(3,4), e1 &		-3.99981	 &	132.07804	 \\
& CH$_{2}$DOH	 &	Methanol &		349.95183610 (8.6E-6) &		 	14(6,8) - 15(5,10), o1 &		-4.60584 &		381.77610 \\	
72 & CH$_{3}$SH v = 0  &	Methyl Mercaptan &		350.00961500 (4.8E-5) &		14( 1) + - 13( 1) +A &		-3.37086 &		131.12515	 \\
73 & CH$_{2}$DOH	 &	Methanol &		350.02734890 (8.0E-6) &		 	6(4,3) - 6(3,3), e1 &		-4.04307 &		117.08865	 \\
& CH$_{2}$DOH	 &	Methanol &		350.02776910 (8.0E-6) &		 	6(4,2) - 6(3,4), e1 &		-4.04307 &		117.08867	 \\
\hline
\multicolumn{7}{c}{spw8}\\
\hline
74 & CH$_{3}$CHO v = 0	 &	Acetaldehyde &		350.08080620 (7.9E-6) &	 	6(3, 4) - 5(2, 4), A &		-3.90056 &		39.71286	 \\
75 & CH$_{2}$DOH	 &	Methanol &		350.09024240 (9.4E-6) &		5(4,2) - 5(3,2), e1	 &	-4.12044 &		104.24014	 \\
& CH$_{2}$DOH	 &	Methanol &		350.09038410 (9.4E-6) &	 	5(4,1) - 5(3,3), e1	 &	-4.12044	 &	104.24015 \\	
76 & $^{13}$CH$_{3}$OH vt = 0	 &	Methanol &		 	350.10311800 (5.0E-5) &		1( 1, 1)- 0( 0, 0) + + &		-3.48249 &		16.80220	 \\
77 & CH$_{3}$CHO v = 0	 &	Acetaldehyde &		350.13342960 (2.75E-5) &		 	18(3, 15) - 17(3, 14), E &		-2.82552 &		179.20917	 \\
78 & CH$_{3}$CHO v = 0	 &	Acetaldehyde &		350.13438160 (2.75E-5) &		 	18(3, 15) - 17(3, 14), E &		-2.82596	 &	179.17727 \\	
\hline
\multicolumn{7}{c}{spw9}\\
\hline
79 & $^{13}$CH$_{3}$OH vt = 0	 &	Methanol &		 	350.42158500 (5.0E-5)  &	8 ( 1 , 7)- 7 ( 2 , 5)	 &	-4.15313 &		102.61554	 \\
80 & CH$_{3}$OCHO v=0	 &	Methyl Formate &		 	350.44225000 (0.0001) &		28( 8,21)-27( 8,20) E	 &	-3.21991	 &	283.90675 \\
81 & CH$_{3}$CHO v = 0	 &	Acetaldehyde &		350.44577770 (2.74E-5) &		 	18(1, 17) - 17(1, 16), E	 &	-2.81490	 &	163.41952 \\
82 & CH$_{3}$OCHO v=0	 &	Methyl Formate &		 	350.45758000 (5.0E-5) &		28( 8,21)-27( 8,20) A	 &	-3.21978	 &	283.90950 \\
\hline
\multicolumn{7}{l}{Note. Species with the same numbering are observed as a blended line.}
\end{longtable}

\clearpage
\renewcommand{\figurename}{{\bf Supplementary Figure }}
% extended data figure
\begin{figure*}
\includegraphics[height=6cm]{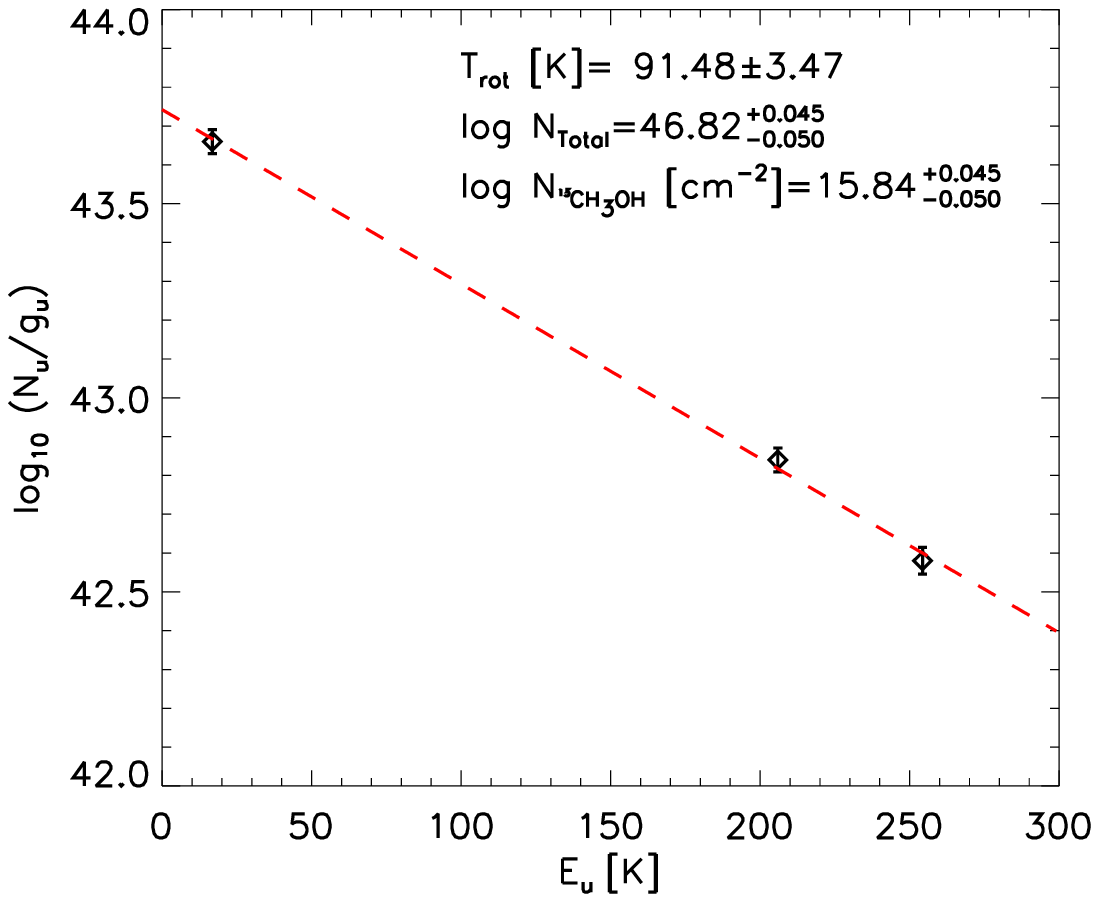}
\includegraphics[height=6cm]{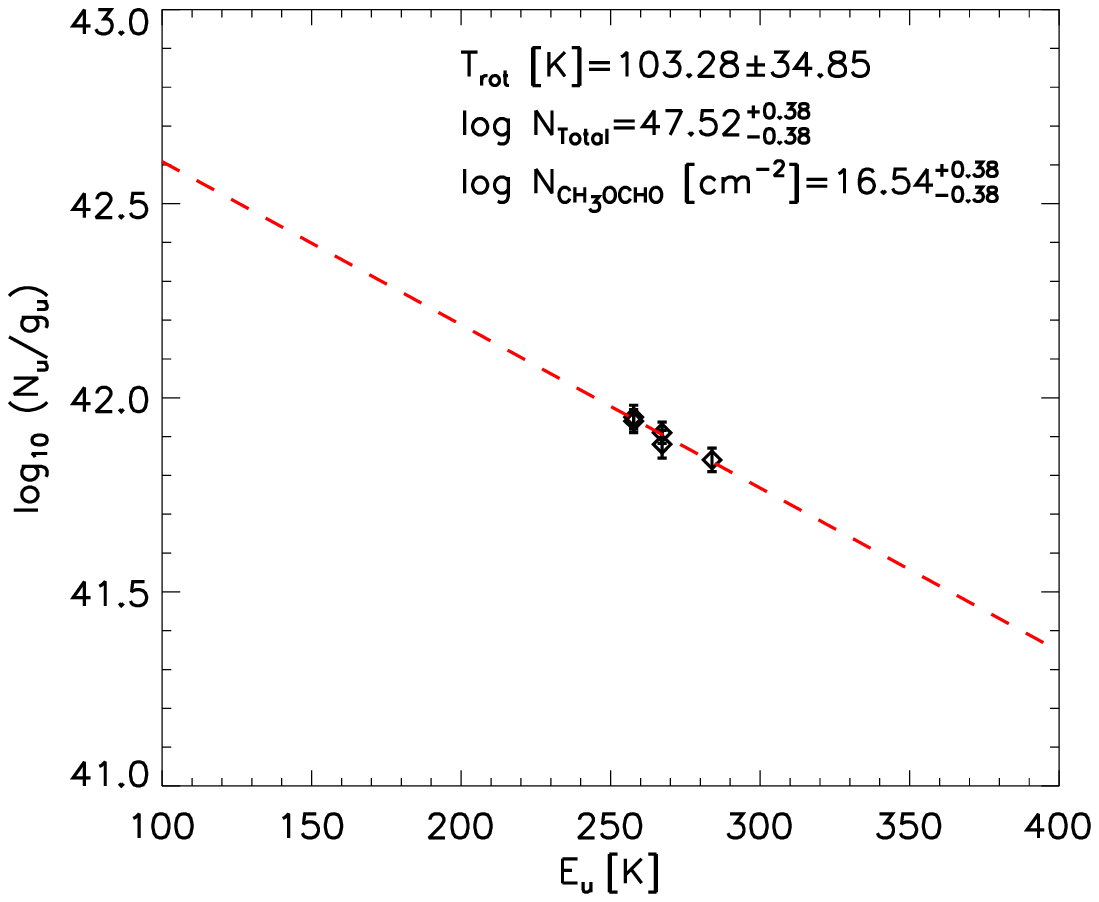}
	\caption{ Excitation Diagrams of $^{13}$CH$_3$OH and CH$_3$OCHO. 
	We plot the log of total number of molecules per degenerate sublevel ({\rm $N_u/g_u$}) versus the energy in the upper state ({\rm $E_u$}). In order to estimate the column density, we applied the aperture size (0.6\arcsec) used for the extraction of lines.
The column densities derived by the excitation diagrams are smaller than the values derived from the XCLASS fitting probably because the dust continuum emission is optically thick. The error bars on the data points and the temperature and column density errors in the legend indicate the 1 $\sigma$ error.}
% 1d-4 /560/ 4d (opr = 3:1) ~ 4.4e-8
\label{fig:fig_s1}
\end{figure*}

\begin{figure*}
\includegraphics[height=7cm]{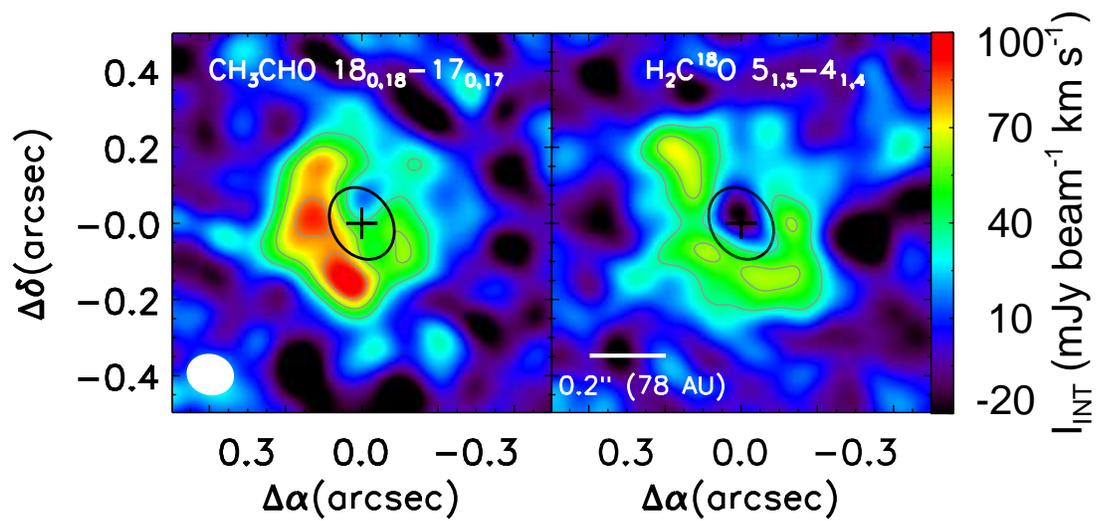}
\caption{ The same images as Figure 3 except for the CH$_3$CHO 18$_{18}$-17$_{17}$ (left) and H$_2$C$^{18}$O 5$_5$-4$_4$ (right) lines. These line emission images are also missing flux within the water snow line. 
  }
\label{fig:fig_s2}

\end{figure*}
\begin{figure*}
\includegraphics[height=7cm]{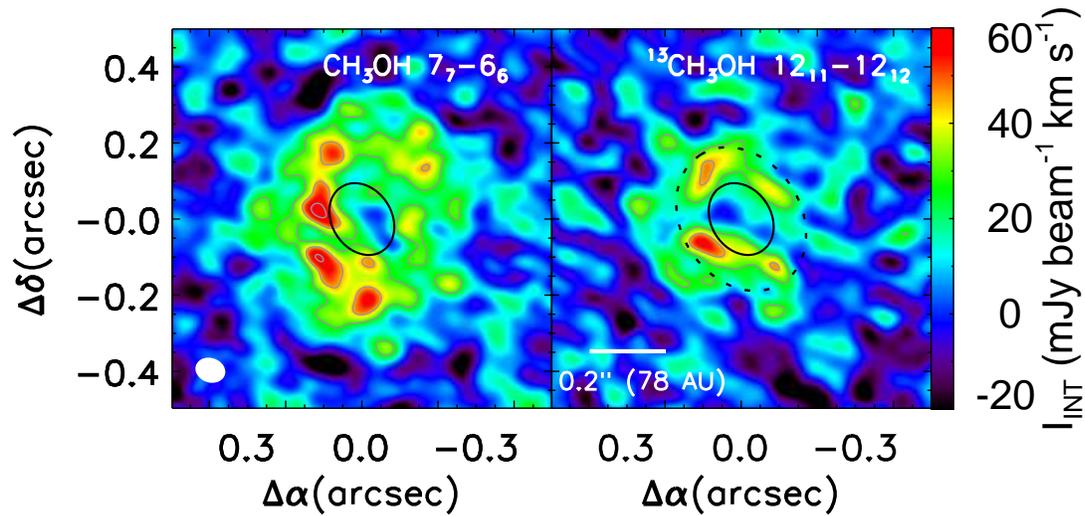}
\caption{ Integrated intensity maps of CH$_3$OH 7$_7$-6$_6$ (left) and $^{13}$CH$_3$OH 12$_{11}$-12$_{12}$ (right) from the high resolution observation.
  A uv-taper of 2000 K$\lambda$ was applied to make images with signal-to-noise ratios better than the original images and a resolution ($\sim 0.07\arcsec$), higher than those in Figure 2(a).
  The lowest contour and subsequent contour step are 5  and 3 $\sigma$ ($\sigma=5.5$ mJy beam$^{-1}$ km s$^{-1}$).
%  The identified water snow line\cite{Cieza2016} is marked with the black solid ellipses. The dotted ellipse describes the projected $0.2\arcsec$ radius.
  The identified water snow line$^4$ is marked with the black solid ellipses. The dotted ellipse describes the projected $0.2\arcsec$ radius.
The synthesized beam is plotted in the left bottom corner. 
  }
\label{fig:fig_s3}
\end{figure*}

\begin{figure*}
\includegraphics[height=11cm]{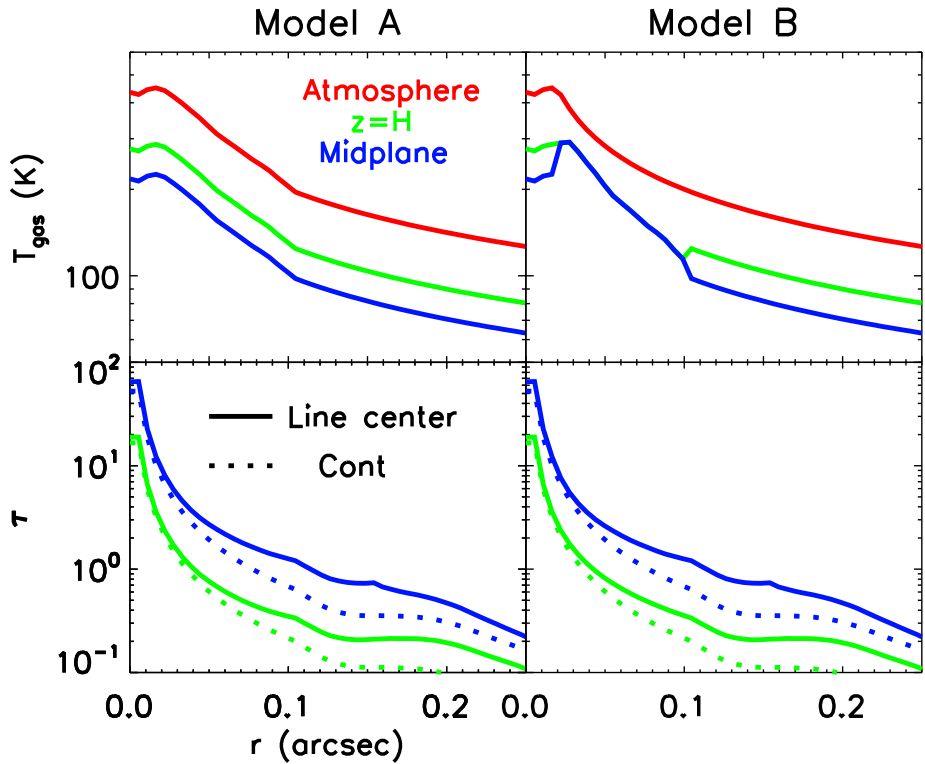}
\caption{ Radial distribution of temperature (upper) and optical depths (lower) for models.
%The grey model (leftmost) fits the Band 6 continuum image. 
%The solid black line shows the radial distribution of the averaged dust temperature ($T_{\rm dust}$) while the dashed black line
%indicate the surface temperature ($T_{\rm surface}$) of the irradiated disk, which follows a profile of $1/\sqrt{r}$. The leftmost lower panel describes the dust continuum optical depth profile at Band 6.
Upper panels show the temperature profiles used for the irradiated disk (Model A) and the combination of the irradiated disk and the heated midplane by a burst accretion (Model B).
The temperature profiles at the atmosphere, z $=H$, and the midplane are presented with the red, green, and blue lines, respectively.
In the lower panels, the solid blue and green lines show the optical depths at the line center of $^{13}$CH$_3$OH 12$_{11}$-11$_{12}$ at the midplane and z $=H$, respectively. The dotted lines represent the Band 7 continuum optical depths at the same heights.
At very small radii, the continuum optical depth dominates the line optical depth.
}
\label{fig:fig_s4}
\end{figure*}

\begin{figure*}
\includegraphics[height=13cm]{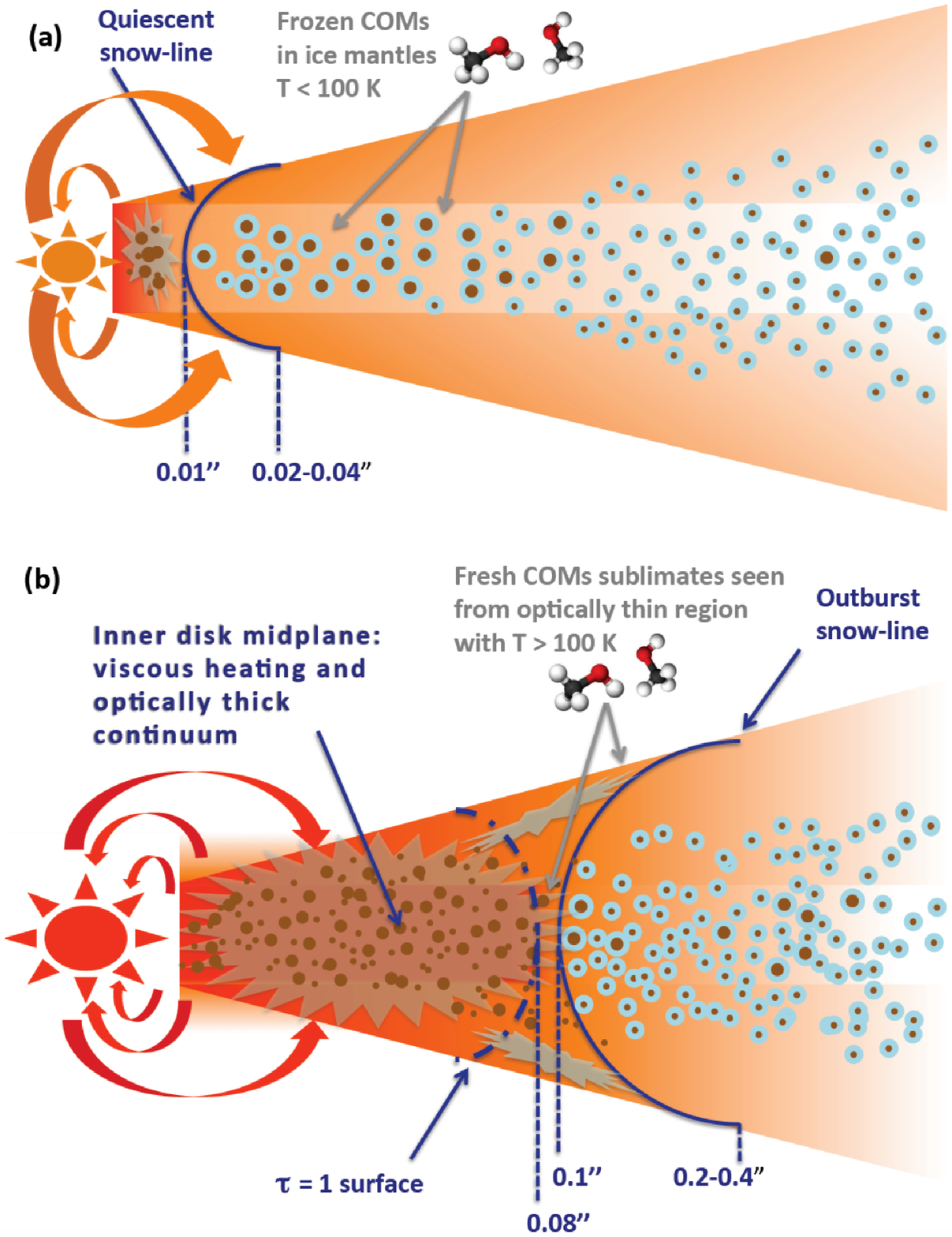}
\caption{ Sketch for the observed phenomenon. The outburst moves the snow line outward. The irradiation from the central protostar heats the disk surface with the positive temperature gradient at most radii to make a 2-D snow surface, i.e., the snow line extends further out at the disk surface. On the other hand, at very small radii, the accretion heats the disk midplane to result in a negative temperature gradient. In the inner disk, the dust continuum opacity is too high to detect the COMs lines in spite of their sublimation. Around the water snow line, the continuum opacity becomes low enough for COMs lines to be detected. The dashed-dotted line indicates the surface with the continuum optical depth of 1.}
% 1d-4 /560/ 4d (opr = 3:1) ~ 4.4e-8
\label{fig:fig4h2o}
\end{figure*}

\end{document}